\documentclass[12pt]{article}
\usepackage{amsmath,amsthm,amsfonts,amssymb}
\def\isom{\cong}

\def\lan{\langle}
\def\ran{\rangle}
\def\Ad{{\mathrm {Ad}}}
\def\Aut{{\mathrm {Aut}}}
\def\End{{\mathrm {End}}}
\def\Hom{{\mathrm {Hom}}}
\def\id{{\mathrm {id}}}
\def\o{{\mathrm {opp}}}
\def\op{{\mathrm {opp}}}
\def\Tr{{\mathrm {Tr}}}
\def\Vir{{\mathrm {Vir}}}

\def\a{\alpha}
\def\be{\beta}
\def\de{\delta}
\def\e{\varepsilon}
\def\f{\varphi}
\def\epsilon{\varepsilon}
\def\ga{\gamma}

\def\la{\lambda}

\def\phi{\varphi}
\def\si{\sigma}
\def\th{\theta}
\def\om{\omega}
\def\Om{\Omega}

\newtheorem{theorem}{Theorem}[section]
\newtheorem{lemma}[theorem]{Lemma}

\newtheorem{corollary}[theorem]{Corollary}
\newtheorem{definition}[theorem]{Definition}

\newtheorem{proposition}[theorem]{Proposition}
\newtheorem{remark}[theorem]{Remark}
\newtheorem{example}[theorem]{Example}

\def\setminus{\smallsetminus}
\def\Diff{{\mathrm {Diff}}}

\def\ov{\overline}
\def\A{{\cal A}}
\def\B{{\cal B}}
\def\cC{{\cal C}}

\def\O{{\cal O}}
\def\K{{\cal K}}
\def\G{{\mathsf G}}

\def\L{{\cal L}}

\def\N{{\cal N}}
\def\M{{\cal M}}
\def\E{{\cal E}}
\def\H{{\cal H}}
\def\T{{\cal T}}
\def\U{{\cal U}}
\def\Z{{\mathbb Z}}
\def\R{{\mathbb R}}
\def\C{{\mathbb C}}

\title{{\bf Classification of Two-dimensional\\
Local Conformal Nets with $c<1$\\
and 2-cohomology Vanishing\\
for Tensor Categories}}

\author{
{\sc Yasuyuki Kawahigashi}\footnote{Supported in part by JSPS.}\\
Department of Mathematical Sciences\\
University of Tokyo, Komaba, Tokyo, 153-8914, Japan\\
e-mail: {\tt yasuyuki@ms.u-tokyo.ac.jp}\\
\vphantom{X}\\
{\sc Roberto Longo}\footnote{Supported in part by GNAMPA and MIUR.}\\
Dipartimento di Matematica\\
Universit\`a di Roma ``Tor Vergata''\\
Via della Ricerca Scientifica, 1, I-00133 Roma, Italy\\
e-mail: {\tt longo@mat.uniroma2.it}}
\begin{document}
\date{April 14, 2003}
\maketitle

\begin{abstract}
We classify two-dimensional local conformal nets with 
parity symmetry and central charge less than 1, up to isomorphism.
The maximal ones are in a bijective correspondence
with the pairs of $A$-$D$-$E$ Dynkin diagrams with
the difference of their Coxeter numbers equal to 1.   In our
previous classification of one-dimensional local conformal nets,
Dynkin diagrams $D_{2n+1}$ and $E_7$ do not appear, but now
they do appear in this classification of two-dimensional
local conformal nets.  Such 
nets are also characterized as two-dimensional local conformal nets
with $\mu$-index equal to 1 and central charge less than 1.
Our main tool, in addition to our previous classification results for
one-dimensional nets, is 2-cohomology vanishing for
certain tensor categories related to the Virasoro tensor categories
with central charge less than 1.
\end{abstract}
%\newpage

\section{Introduction}

The subject of Conformal Quantum Field Theory is particularly 
interesting in two spacetime dimensions and has indeed been 
intensively studied in 
the last two decades with important motivations from Physics (see 
e.g. \cite{DMS}) and Mathematics (see e.g. \cite{EK}). 
Basically the richness of structure is due to the fact that conformal 
group (with respect to the Minkowskian metric) is infinite 
dimensional in $1+1$ dimensions.

Already at the early stage of investigation, it was realized that such 
infinite dimensional symmetry group puts rigid constrains on structure 
and the problem of classification of all models was posed and 
considered as a major aim. Indeed many important results in 
this direction were obtained, in particular the central charge 
$c>0$, an intrinsic quantum label associated with each model, was 
shown to split in a discrete range $c<1$ and a continuos one 
$c\geq 1$, see \cite{BPZ,FQS,GO} refs in \cite{GO}.

The main purpose of this paper is to achieve a complete classification 
of the two-dimensional conformal models in the discrete series. In order 
to formulate such a statement in a precise manner, we need to 
explain our setting.  

The essential, intrinsic structure of a given model is described 
by a net $\A$ 
on the two-dimensional Minkowski spacetime $\M$ \cite{H}. With each double 
cone $\O$ (an open region which is the intersection of the past of one 
point and the future of a second point) one associates the von Neumann 
algebra $\A(\O)$ generated by the observables localized in $\O$ (say 
smeared fields integrated with test functions with support in $\O$). 

The net $\A: \O\mapsto \A(\O)$ is then local and covariant with 
respect to the conformal group. One may restrict $\A$ to the two 
light rays $x\pm t=0$ and obtain two local conformal nets $\A_{\pm}$ 
on $\mathbb R$, hence on its one point compactification $S^1$. So we 
have an irreducible two-dimensional subnet
\[
\B(\O)\equiv \A_+(I_+)\otimes\A_-(I_-)\subset \A(\O)\ ,
\]
where $\O=I_+\times I_-$ is the double cone associated with the 
intervals $I_{\pm}$ of the light rays.  The structure of $\A$, thus 
the classification of local conformal nets, splits in the following 
two points: 
\begin{itemize}
\item The classification of local conformal nets on $S^1$.
\item The classification of irreducible local extension of chiral 
conformal nets.
\end{itemize}
Here a chiral net is a net that splits in tensor product of two 
one-dimensional nets on the light rays.  Now the conformal group of 
$\M$ is $\Diff(S^1)\times \Diff(S^1)$ 
 \footnote{More precisely $\Diff(S^1)\times \Diff(S^1)$ is the 
conformal group of the Minkowskian torus $S^1\times S^1$, the conformal 
completion of $\M=\mathbb R\times\mathbb R$ (light ray 
decomposition), and the covariance group is a central extension of
$\Diff(S^1)\times \Diff(S^1)$, see Sect. \ref{Sect2}.} 
thus, restricting the projective unitary covariance representation to 
the two copies of $\Diff(S^1)$, we get Virasoro nets 
$\Vir_{c_{\pm}}\subset \A_{\pm}$ with central charge $c_{\pm}$.  If 
there is a parity symmetry, then $c_+ = c_-$, so we may talk of the 
central charge $c\equiv c_{\pm}$ of $\A$.  If $c<1$, it turns out by 
that $\A$ is completely rational \cite{KLM} and the subnet 
$\Vir_c\otimes\Vir_c\subset\A$ has finite Jones index, where
$\Vir_c\otimes\Vir_c(\O)\equiv\Vir_c(I_+)\otimes\Vir_c(I_-)$. 

The classification of two-dimensional local conformal nets with central 
charge $c<1$ and parity symmetry thus splits in the following two points:
\begin{itemize}
\item[$(a)$] The classification of Virasoro nets $\Vir_c$ on $S^1$ with $c<1$.
\item [$(b)$] The classification of irreducible local extensions with finite 
Jones index of the two-dimensional Virasoro net $\Vir_c\otimes\Vir_c$.
\end{itemize}
Point $(a)$ has been completely achieved in our recent work \cite{KL}. The 
Virasoro nets on $S^1$ with central charge less than one are in bijective 
correspondence with the pairs of $A$-$D_{2n}$-$E_{6,8}$ Dynkin diagrams 
such that 
the difference of their Coxeter numbers is equal to 1. Among other 
important aspects of this classification, we mention here the 
occurrence of nets 
that are not realized as coset models, in contrast to a long standing 
expectation.  (See Remarks after Theorem 7 of \cite{Ks} on 
tihs point.  Also, Carpi and Xu recently made a progress on 
classification for the case $c=1$ in \cite{C}, \cite{X3},
respectively.)

The aim of this paper is to pursue point $(b)$.  We shall obtain a 
complete classification of the two-dimensional local conformal 
nets (with parity) with central charge in the discrete series. To this 
end we first classify the maximal nets in this class. Maximality here 
means that the net does not admit any irreducible local conformal net 
extension.  Maximality will turn out to be also equivalent to the 
triviality of the superselection structure or to $\mu$-index equal to 
one, that is Haag duality for disconnected union of finitely many 
double cones.

It is clear at this point that our methods mainly concern Operator 
Algebras, in particular Subfactor Theory, see \cite{T}.  Indeed this 
was already the case in our previous one-dimensional classification 
\cite{KL}.  The use of von Neumann algebras not only provides a clear 
formulation of the problem, but also suggests the path to follow in the 
analysis.

Our strategy is the following. The dual canonical endomorphism of 
$\Vir_c\otimes \Vir_c\subset\A$ decomposes as 
\begin{equation}\label{th}
\th = \bigoplus_{ij} Z_{ij} \rho_i\otimes\bar\rho_j
\end{equation}
(i.e. the above is the restriction to $\Vir_c\otimes \Vir_c$ of the 
vacuum representation of $\A$), where $\{\rho_i\}_i$ are representatives
of unitary equivalence classes of irreducible DHR endomorphisms 
of the net $\Vir_c$.

Since $\mu_\A=1$ it turns out, by using the results in \cite{KLM}, 
that the matrix $Z$ is a modular invariant for the tensor category of 
representations of the Virasoro net $\Vir_c$ \cite{Mu}, and such 
modular invariants have been classified by Cappelli-Itzykson-Zuber 
\cite{CIZ}.

We shall show that this map $\A\mapsto Z$ sets up a bijective 
correspondence between the the set of isomorphism classes of two-dimensional 
maximal local conformal nets with parity and central charge less than 
one on one hand and the list of Cappelli-Itzykson-Zuber modular 
invariant on the other hand.

We first prove that the correspondence $\A\mapsto Z$ is surjective.
Indeed, by our previous work \cite{KL}, $Z$ can be realized
by $\a$-induction as in \cite{BEK1} for extensions of the Virasoro nets.
(See \cite{LR,X1,BE,BEK2,BEK3,BE4} for more on $\a$-induction.)
Then Rehren's results in \cite{R3} imply that $\th$ defined as above 
(\ref{th}) is the canonical endomorphism associated  with a natural
$Q$-system, and we have a corresponding local extension 
$\A$ of $\Vir_c\otimes \Vir_c$ 
and this produces the matrix $Z$ in the above correspondence.

To show the injectivity of the correspondence
note that, due to the work of Rehren \cite{R2}, we have an inclusion
\[
\Vir_c(I_+)\otimes\Vir_c(I_-)\subset 
\A_+(I_+)\otimes\A_-(I_-)\subset\A(\O)
\]
where $\A_+\otimes\A_-$ is the maximal chiral subnet.
By assumption, $\A_+$ and $\A_-$ are isomorphic with central
charge $c<1$, thus they are in the discrete series classified in \cite{KL}.
Moreover $Z$ determines uniquely the isomorphism class of $\A_{\pm}$ 
and an isomorphism $\pi$ from a fusion rule of $\A_+$ onto that
of $\A_-$ so that the dual canonical endomorphism
$\la$ on $\A_+\otimes\A_-$ decomposes as
\begin{equation}\label{theta}
\la=\bigoplus_{i} \a_i \otimes\bar\a_{\pi(i)},
\end{equation}
where $\{\a_i\}_i$ is a system of irreducible DHR 
endomorphisms of $\A_+ = \A_-$. 

If $Z$ is a modular invariant of type I, the map $\pi$ is trivial, so 
the dual canonical endomorphism has the same form of the Longo-Rehren 
endomorphism \cite{LR}.  Thus the classification is reduced to 
classification of $Q$-systems in the sense of \cite{L2} having the 
canonical endomorphism of the form given by eq.  (\ref{th}).  This 
type of classification of $Q$-systems, up to unitary equivalences, was 
studied by Izumi-Kosaki \cite{IK} as a subfactor analogue of 
2-cohomology of (finite) groups.  In our setting, we now have a 
2-cohomology \emph{group} of a tensor category, while the 2-cohomology 
of Izumi-Kosaki does not have a group structure in general.  The group 
operation comes from a natural composition of 2-cocycles.  Then the 
crucial point in our analysis is the vanishing of this 2-cohomology 
for certain tensor category as we will explain below, and this 
vanishing implies that the dual $Q$-system for the inclusion 
$\A_+\otimes \A_-\subset \A$ has a standard dual 
canonical endomorphism as in the Longo-Rehren $Q$-system \cite{LR}, 
namely $\A_+\otimes \A_-\subset \A$ is the ``quantum 
double'' inclusion constructed in \cite{LR}.  At this point, as we 
know the isomorphism class of $\A_{\pm}$ by our previous 
classification \cite{KL}, it follows that the isomorphism class of 
$\A$ is determined by $Z$.

If the modular invariant is of type II, then $\pi$ gives
a non-trivial fusion rule automorphism, however $\pi$ is
actually associated with an automorphism of the tensor category
acting non-trivially on irreducible objects \cite{BE4}.
We may then extend our arguments of 2-cohomology vanishing
and deal also with this case.  It turns out that the
automorphism $\pi$ is an automorphism of a \emph{braided}
tensor category.

We thus arrive at the following classification: the maximal local 
two-dimensional conformal nets with $c<1$ and parity symmetry are in a 
bijective correspondence with the pairs of the $A$-$D$-$E$ Dynkin 
diagrams such that the difference of their Coxeter numbers is equal to 
$1$, namely $Z$ is a modular invariant listed in Table \ref{CIZ-mod}.  
Note that Dynkin diagrams of type $D_{2n+1}$ and $E_7$ do appear in 
the list of present classification of two-dimensional conformal nets, 
but they were absent in the one-dimensional classification list 
\cite{KL}.

Now, as we shall see, two-dimensional local conformal net $\B$ 
in the discrete series is a finite-index subnet of a maximal 
local conformal net $\A$. Moreover $\A$ and $\B$ have the same 
two-dimensional Virasoro subnet.  Using this, we then obtain
the classification of all local two-dimensional conformal nets 
with central charge less $c<1$. The non-maximal ones correspond bijectively
to the pairs $(\T,\a)$ where $\T$ is a proper sub-tensor category of 
the representation tensor category of $\Vir_c$ and $\a$ is an 
automorphism of $\T$. There are at most two automorphisms, thus two 
possible nets for a given $\T$. The complete list is given in Table 
\ref{tab-sub-cat}.

As we have mentioned, a crucial point in our analysis is to show the 
uniqueness up to equivalence of the $Q$-system associated with the 
canonical endomorphism of the form (\ref{theta}) in our cases. To this 
end we consider a cohomology associated with a representation tensor 
category that we have to show to vanish in our case.   Note that
our 2-cohomology groups are generalization of the usual 2-cohomology
groups of finite groups, so they certainly do not vanish in general.

Before concluding this introduction we make explicit that our 
classification applies as well to the local conformal nets with 
central charge less than one on other two-dimensional spacetimes.  
Indeed if $\N$ is two-dimensional spacetime that is conformally 
equivalent to $\M$, namely conformally diffeomorphic to a subregion on 
the Einstein cylinder $S^1\times\mathbb R$, we may then consider the 
local conformal nets on $\N$ that satisfy the double cone KMS property.  
These nets are in one-to-one correspondence with the local conformal 
nets on Minkowski spacetime $\M$, see \cite{GL2}, and so one 
immediately reads off our classification in these different contexts.  An 
important case where this applies is represented by the 
two-dimensional de Sitter spacetime.
 
\section{Two-dimensional completely rational nets and central charge}
\label{Sect2}

Let $\M$ be the two-dimensional Minkowski spacetime, 
namely $\mathbb R^2$ equipped with the metric 
$\text{d}t^2 - \text{d}x^2$. We shall also use the light ray coordinates
$\xi_{\pm}\equiv t \pm x$. We have the decomposition 
$\M=\L_+\times\L_-$ where $\L_{\pm}=\{\xi:\xi_{\pm}=0\}$ are the two 
light ray lines. A \emph{double cone} $\O$ is a non-empty open subset of
of $\M$ of the form $\O=I_+\times I_-$ with $I_{\pm}\subset\L_{\pm}$  
bounded intervals; we denote by $\K$ the set of double cones.

The M\"{o}bius group $PSL(2,\mathbb R)$ acts on $\mathbb 
R\cup\{\infty\}$ by linear fractional transformations, hence this 
action restricts to a local action on $\mathbb R$ (see e.g.  
\cite{BGL}), in particular if $F\subset\mathbb R$ has compact closure 
there exists a connected neighborhood $\U$ of the identity in 
$PSL(2,\mathbb R)$ such that $gF\subset\R$ for all $g\in\U$.  It is 
convenient to regard this as a local action on $\R$ of the universal 
covering group $\ov{PSL}(2,\mathbb R)$ of $PSL(2,\mathbb R)$.  We then 
have a local (product) action of 
$\ov{PSL}(2,\mathbb R)\times\ov{PSL}(2,\mathbb R)$ on 
$\M=\L_+\times\L_-$. 
Clearly $\ov{PSL}(2,\mathbb R)\times\ov{PSL}(2,\mathbb R)$
acts by pointwise rescaling the 
metric $\text{d}\xi_+\text{d}\xi_-$, i.e. by conformal transformations.

A \emph{local M\"{o}bius covariant net} $\A$ on $\M$ is a map
\[
\A:\O\in\K\mapsto\A(\O)
\]
where the $\A(\O)$'s are von Neumann algebras on a fixed Hilbert 
space $\H$, with the following properties:
\begin{itemize}

\item \emph{Isotony.} $\O_1\subset\O_2\implies 
\A(\O_1)\subset\A(\O_2)$.

\item \emph{Locality.} If $\O_1$ and $\O_2$ are spacelike separated 
then $\A(\O_1)$ and $\A(\O_2)$ commute elementwise (two points $\xi_1$
and $\xi_2$ are spacelike if $(\xi_1 -\xi_2)_+(\xi_1 -\xi_2)_- <0$).

\item \emph{M\"{o}bius covariance.} There exists a unitary representation 
$U$ of $\ov{PSL}(2,\mathbb R)\times\ov{PSL}(2,\mathbb R)$ 
on $\H$ such that, for every double cone $\O\in\K$,
\[
U(g)\A(\O)U(g)^{-1} = \A(g\O),\quad g\in\U,
\]
with $\U\subset\ov{PSL}(2,\mathbb R)\times\ov{PSL}(2,\mathbb R)$ 
any connected neighborhood of the identity 
such that $g\O\subset\M$ for all $g\in\U$.

\item \emph{Vacuum vector.} There exists a unit $U$-invariant vector 
$\Omega$, cyclic the $\bigcup_{\O\in\K}\A(\O)$.

\item \emph{Positive energy.} The one-parameter unitary subgroup of $U$ 
corresponding to time translations has positive generator.

\end{itemize}
The $2$-torus $S^1\times S^1$ is a conformal completion of 
$\M=\L_+\times\L_-$ in the sense that $\M$ is conformally diffeomorphic to a 
dense open subregion of $S^1\times S^1$ and the local action of 
$PSL(2,\mathbb R)\times PSL(2,\mathbb R)$ on $\M$ extends to a global 
conformal action on $S^1\times S^1$.

But in general the net $\A$ does not extend to a M\"{o}bius covariant 
net on $S^1\times S^1$; this is related to the failure of 
timelike commutativity (note that a chiral net, i.e.  the tensor 
product of two local nets on $S^1$, would extend), indeed we have a 
covariant unitary representation of 
$\ov{PSL}(2,\mathbb R)\times\ov{PSL}(2,\mathbb R)$ 
and not of $PSL(2,\mathbb R)\times PSL(2,\mathbb R)$.

Let however $\G$ be the quotient of $\ov{PSL}(2,\mathbb 
R)\times\ov{PSL}(2,\mathbb R)$ modulo the relation 
$(r_{2\pi},r_{-2\pi})=(\text{id},\text{id})$ 
(spatial $2\pi$-rotation is the identity). 
\begin{proposition}\label{factor}
The representation $U$ of $\ov{PSL}(2,\mathbb R)\times\ov{PSL}(2,\mathbb R)$ 
factors through a representation of $\G$.
\end{proposition}
The above proposition holds as a consequence of spacelike locality, it 
is a particular case of the conformal spin-statistics theorem and can 
be proved as in \cite{GL1}.

Because of the above Prop.  \ref{factor}, $\A$ does extend to a local 
$\G$-covariant net on the Einstein cylinder $\E=\R\times S^1$, the 
cover of the $2$-torus obtained by lifting the time coordinate from 
$S^1$ to $\mathbb R$.
 
Explicitly, $\M$ is conformally equivalent to a double cone 
$\O_{\M}$ of $\E$.  By 
parametrizing $\E$ with coordinates $(t',\theta)$, 
$-\infty<t'<\infty$, $-\pi\leq\theta<\pi$, the transformation
\begin{equation}
	\xi_{\pm}=\tan(\tfrac{1}{2}(t'\pm \theta))
	\end{equation}
is a diffeomorphism of the subregion 
$\O_{\M}=\{(t',\th):-\pi < t' \pm\theta < \pi\}\subset\E$ 
with $\M$, which is a conformal map when $\E$ is equipped with 
the metric $\text{d}s^{2}\equiv\text{d}{t'}^{2}-\text{d}\theta^{2}$.

$\G$ acts globally on $\E$ and the net $\A$ extends uniquely to a 
$\G$-covariant net of $\E$ with $U$ the unitary covariant action (see 
\cite{BGL}).  We shall denote by the same symbol $\A$ both the net on 
$\M$ and the extended net on $\E$.

If $\O_1\subset \M$ (or $\O_1\subset \E$) we shall denote by 
$\A(\O_1)$ the von Neumann algebra generated by the $\A(\O)$'s as $\O$ 
varies in the double cones contained in $\O_1$.  If $\O\in\K$ we shall 
denote by $\Lambda_{\O}$ the one-parameter subgroup of $\G$ defined as 
follows: $\Lambda_{\O}=g\Lambda_{W}g^{-1}$ if $W$ is a wedge, 
$\Lambda_{W}$ is the boost one-parameter group associated with $W$, 
and $g\O=W$ with $g\in PSL(2,\mathbb R)\times PSL(2,\mathbb R)$, see 
\cite{HL}.

We collect in the next proposition a few basic properties of a local 
M\"{o}bius covariant net. The proof is either in the references or can 
be immediately obtained from those. All the statements also hold true 
(with obvious modifications) in 
any spacetime dimension. We shall use the lattice symbol $\vee$ to 
denote the von Neumann algebra generated.
\begin{proposition}
Let $\A$ be a local M\"{o}bius covariant net on $\M$ as above.
The following hold:
\begin{itemize}
\item[$(i)$] 
\emph{Double cone KMS property.} If $\O\subset \E$ is a double 
cone, then the unitary modular group associated with $(\A(\O),\Om)$ 
has the geometrical meaning $\Delta_{\O}^{it}=U(\Lambda_{\O}(-2\pi 
t))$ {\rm \cite{BGL}}.
\item[$(ii)$] 
\emph{Haag duality on $\E$; wedge duality on $\M$.} 
If $\O\subset \E$ is a double cone then $\A(\O')=\A(\O)'$.  Here $\O'$ 
is the causal complement of $\O$ in $\E$ (note that $\O'$ is still a 
double cone.) In particular $\A(W')=A(W)'$, where $W$ is a wedge in 
$\M$, say $W=(-\infty,a)\times(-\infty,b)$ and $W'$ its causal 
complement in $\M$, thus $W'=(a,\infty)\times(b,\infty)$ 
{\rm \cite{BGL,GL2}}.
\item [$(iii)$]
 \emph{Modular PCT symmetry.} There is a anti-unitary involution 
$\Theta$ on $\H$ such that $\Theta\A(\O)\Theta = \A(-\O)$,\ 
$\Theta U(g)\Theta = U(\th(g))$ and $\theta\Om=\Om$.  
Here $\O$ is any double one in $\E$ and 
$\th$ is the automorphism of $\G$ associated with space and time 
reflection {\rm \cite{BGL}}.
\item[$(iv)$] 
\emph{Additivity.} Let $\O$ be a double 
cone and $\{\O_i\}$ a family of open sets such that $\bigcup_i \O_i$ 
contains the axis of $\O$.  Then $\A(\O)\subset\bigvee_i\A(\O_i)$ 
{\rm \cite{FJ}}.  
\item[$(v)$] 
\emph{Equivalence between rreducibility and uniqueness of the vacuum.} 
$\A$ is irreducible on $\M$ (that is 
$\left(\bigcup_{\O\in\K}\A(\O)\right)''=B(\H)$), iff $\A$ is 
irreducible on $\E$, iff $\Om$ is the unique $U$-invariant vector (up 
to a phase) {\rm \cite{GL1}}.
\item[$(vi)$] 
\emph{Decomposition into irreducibles.} $\A$ has a unique direct 
integral decomposition in terms of local irreducible M\"{o}bius 
covariant nets.  If $\A$ is conformal (see below) then the fibers in 
the decomposition are also conformal {\rm \cite{GL1}}.
\end{itemize}
\end{proposition}
By the above point $(vi)$ we shall always assume our nets to be 
\emph{irreducible}.

Let $\Diff(\R)$ denote the group of positively oriented 
diffeomorphisms of $\R$ that are smooth at infinity (with the 
identification $\mathbb R = S^1\setminus\{\infty\}$, $\Diff(\R)$ is 
the subgroup of $\Diff(S^1)$ of 
orientation preserving diffeomorphisms of $S^1$ 
that fix the point $\infty$). By identifying $\M$ with the double 
cone $\O_\M\subset\E$ as above, we may 
identify elements of $\Diff(\R)\times\Diff(\R)$ with conformal 
diffeomorphisms of $\O_\M$. Such diffeomorphisms uniquely extend (by 
periodicity) to global conformal diffeomorphisms of $\E$. Namely the element 
$(r_{2\pi},\text{id})$ of $\G$ generates a subgroup  of $\G$ 
(isomorphic to $\mathbb Z$)
for which $\O_\M$ is a fundamental domain in $\E$. We may then extend 
an element of $\Diff(\R)\times\Diff(\R)$ 
from $\O_\M$ to all $\E$ by requiring 
commutativity with this $\mathbb Z$-action; this is the unique 
conformal extension to $\E$.

Let $\text{Conf}(\E)$ denote the group of global, orientation preserving 
conformal diffeomorphisms of $\E$. $\text{Conf}(\E)$ is generated by 
$\Diff(\R)\times\Diff(\R)$ and $\G$ (note that $\Diff(\R)\times\Diff(\R)$ 
intersects $\G$ in the ``Poincar\'e-dilation'' subgroup). Indeed if 
$\f\in\text{Conf}(\E)$, then $\f\O_\M$ is a maximal double cone 
of $\E$, namely the causal complement of a point. Thus there exists an 
element $g\in\G$ such that $g\O_\M=\f\O_\M$. Then 
$\psi\equiv g^{-1}\f$ maps $\O_\M$ onto 
$\O_\M$ and so $\psi\in\Diff(\R)\times\Diff(\R)$ and 
$\f=g\cdot\psi$. Note that, by the same argument, any element of 
$\text{Conf}(\E)$ is uniquely the product of an element 
of $\Diff(\R)\times\Diff(\R)$,
a space rotation and time translation on $\E$.

A \emph{local conformal net} $\A$ on $\M$ is a M\"{o}bius covariant 
net such that the unitary representation $U$ of $\G$ extends to a 
projective unitary representation of $\text{Conf}(\E)$ 
(still denoted by $U$) such that so that the extended net on $\E$ is 
covariant. In particular
\[
U(g)\A(\O)U(g)^{-1} = \A(g\O),\quad g\in\U\ ,
\]
if $\U$ is a connected neighborhood of the identity of
$\text{Conf}(\E)$, $\O\in\K$, and $g\O\subset\M$ for all 
$g\in\U$. We further assume that
\[
U(g)XU(g)^{-1} = X,\quad g\in\Diff(\R)\times\Diff(\R)\ ,
\]
if $X\in\A(\O_1)$, $g\in\Diff(\R)\times\Diff(\R)$ and $g$ acts 
identically on $\O_1$. We may check the conformal covariance on $\M$ 
by the local action of $\Diff(\R)\times\Diff(\R)$.

Given a M\"{o}bius covariant net $\A$ on $\M$ and a bounded interval
$I\subset\L_{+}$  we set 
\begin{equation}
\A_{+}(I)\equiv \bigcap_{\O=I\times J}\A(\O)
\end{equation}
(intersection over all intervals $J\subset \L_-$), and analogously 
define $\A_-$. 
By identifying $\L_{\pm}$ with $\R$ we then get two M\"{o}bius covariant 
local nets $\A_{\pm}$ on $\R$, \emph{the chiral components of $\A$}, 
but for the cyclicity of $\Omega$; we shall also denote $\A_{\pm}$ 
by $\A_R$ and $\A_L$. 
By the Reeh-Schlieder theorem the cyclic subspace
$\H_{\pm}\equiv\overline{\A_{\pm}(I)\Omega}$
is independent of the interval $I\subset\L_{\pm}$ and $\A_{\pm}$ 
restricts to a (cyclic) M\"{o}bius covariant local net on $\R$ on the 
Hilbert space $\H_{\pm}$. Since $\Omega$ is separating for every 
$\A(\O)$, $\O\in\K$, the map $X\in\A_{\pm}(I)\mapsto 
X\restriction\H_{\pm}$ is an isomorphism for any interval $I$, so we 
will often identify $\A_{\pm}$ with its restriction to $\H_{\pm}$.
\begin{proposition}
Let $\A$ be a M\"{o}bius covariant (resp.  conformal) net on $\M$.  
Setting $\A_0(\O)\equiv \A_+(I_+)\vee\A_-(I_-)$, $\O=I_+\times I_-$, 
then $\A_0$ is a M\"{o}bius covariant (resp.  conformal) subnet of 
$\A$, there exists a consistent family of vacuum preserving 
conditional expectations $\epsilon_\O:\A(O)\to\A_0(\O)$ and the 
natural isomorphism from the product $\A_+(I_+)\cdot\A_-(I_-)$ to the 
algebraic tensor product $\A_+(I_+)\odot\A_-(I_-)$ 
extends to a normal isomorphism between $\A_+(I_+)\vee\A_-(I_-)$ and 
$\A_+(I_+)\otimes\A_-(I_-)$.
\end{proposition}
Thus we may identify $\H_+\otimes\H_-$ with 
$\H_0\equiv\overline{\A_0(\O)\Omega}$ and $\A_+(I_+)\otimes\A_-(I_-)$ with
$\A_0(\O)$.

It is easy to see that $\A_0$ is the unique maximal chiral subnet 
of $\A$, namely it coincides with the subnet 
$\A_L^{\max}\otimes\A_R^{\max}$ in Rehren's work \cite{R2,R3}. 
That is to say 
$\A_L^{\max}(\O)\otimes 1 =\A(\O)\cap U
\left(\{\text{id}\}\times PSL(2,\mathbb R)\right)'$ and similarly for 
$\A_R^{\max}$. Indeed 
$\A_L^{\max}\otimes\A_L^{\text{max}}$, being chiral, is clearly contained in 
$\A_0$; on the other hand $\A_+$ commutes with $U\restriction 
\text{id}\times PSL(2,\mathbb R)$ so  $\A_+\subset\A_L^{\max}$ 
and analogously $\A_-\subset\A_R^{\max}$.

Now suppose that $\A$ is conformal. We have
\begin{proposition}
If $\A$ is conformal then 
$\A_0\equiv\A_L^{\max}\otimes\A_L^{\max}$ is also 
conformal, moreover $\A_0$ extends to a local 
$\Diff(S^1)\times\Diff(S^1)$-covariant net on the 2-torus, namely 
$\A_{\pm}$ are local conformal nets on $S^1$.
\end{proposition}
Assuming $\A$ to be conformal we set
\begin{gather}
\Vir_+(I)\equiv \big\{ U(g): 
g\in\Diff(I)\times\{\text{id}\} \big\}\,\ I\subset\L_+\\
\Vir_-(I)\equiv \big\{ U(g): 
g\in\{\text{id}\}\times\Diff(I) \big\},\ I\subset\L_- \\
\Vir(\O)\equiv \Vir_+(I_+)\otimes\Vir_-(I_-),\ I_{\pm}\subset\L_{\pm}
\end{gather}
\begin{proposition}
$\Vir_{\pm}(I)\subset \A_{\pm}(I)$, $I\subset\L_{\pm}$, and
$\Vir_+(I_+)\vee\Vir_-(I_-)$ is naturally isomorphic to
$\Vir_+(I_+)\otimes\Vir_-(I_-)$, $I_{\pm}\subset\L_{\pm}$.

$\Vir_{\pm}$ is the restriction to $\L_{\pm}$ of the Virasoro subnet 
of $\A_{\pm}$.
\end{proposition}

In this paper we shall use only the a priori weaker form of conformal 
covariance given by the above proposition. Indeed we shall just need that 
$\A_{\pm}$ are conformal nets on $S^{1}$, with central charge 
less than one.

\subsection{Complete rationality}

Let $\A$ be a local conformal net on the two-dimensional Minkowski 
spacetime $\M$. We shall say that $\A$ is \emph{completely rational} 
if the following three conditions hold:
\begin{itemize}
\item[$a)$] \emph{Haag duality on $\M$.} For any double cone $\O$ we 
have $\A(\O)=\A(\O')'$. Here $\O'$ is the causal complement of $\O$ 
in $\M$
\item[$b)$] \emph{Split property.} If $\O_1 , \O_2 \in\K$ and the 
closure of $\bar\O_1$ of $\O_1$ is contained in $\O_2$, the natural map 
$\A(\O_1)\cdot\A(\O_2)'\to \A(\O_1)\odot\A(\O_2)'$ extends to 
a normal isomorphism $\A(\O_1)\vee\A(\O_2)'\to \A(\O_1)\otimes\A(\O_2)'$.
\item[$c)$] \emph{Finite $\mu$-index.} Let $E=\O_1\cup\O_2\subset\M$ 
be the union of two double cones $\O_1 , \O_2$ such that $\bar\O_1$ 
and $\bar\O_2$ are spacelike separated. Then the Jones index 
$[\A(E')':\A(E)]$ is finite. This index is denoted by $\mu_\A$, 
\emph{the $\mu$-index of $\A$}.
\end{itemize}

The notion of complete rationality has been introduced and studied in 
\cite{KLM} for a local net $\cC$ on $\mathbb R$. If $\cC$ is conformal, 
the definition of complete rationality strictly parallels the above 
one in the two-dimensional case. In general, the above 
(one-dimensional version) of the above three conditions must be 
supplemented by the following two conditions
\begin{itemize}
\item[$d)$] 
\emph{Strong additivity.} If $I_1, I_2\subset\mathbb R$ 
are open intervals and $I$ is the interior of $\overline{I_1\cup 
I_2}$, then $\cC(I)=\cC(I_1)\vee\cC(I_2)$. 
\item [$e)$] 
\emph{Modular PCT symmetry.} 
There is a vector $\Om$, cyclic and separating for all 
the $\cC(I)$'s, such that if $a\in\mathbb R$ the modular conjugation 
$J$ of $(\cC(a,\infty),\Om)$ satisfies $J\cC(I)J = \cC(I+2a)$, for all 
intervals $I$.
\end{itemize}

If $\cC$ is conformal, then $d)$ and $e)$ follows from $a),b),c)$. In 
any case all conditions $a)$ to $e)$ have the strong consequences on 
the structure of $\A$ \cite{KLM}. In particular
\[
\mu_{\cC} = \sum_i d(\rho_i)^2
\]
where the $\rho_i$ form a system of irreducible sectors of $\cC$.

Returning to the two-dimensional local conformal net $\A$, consider 
the time-zero net
\[
\cC(I)\equiv\A(\O),
\]
where $I$ is an interval of the $t=0$ line in $\M$ and $\O=I''$ is the 
double cone with basis $I$. Note that $\cC$ is local but \emph{not} conformal 
(positivity of energy does not hold). However $\cC$ inherits all 
properties from $a)$ to $e)$ from $\A$. Thus we may define $\A$ to be 
completely rational by requiring $\cC$ to be completely rational. In 
this way all results in \cite{KLM} immediately applies to the 
two-dimensional context.

\section{Modular invariance and $\mu$-index of a net}

Rehren raised a question in \cite[page 351, lines 8--13]{R4}
about modular invariant arising from a decomposition of a 
two-dimensional net and its $\mu$-index.
M\"uger  has then solved the problem affirmatively in \cite{Mu}.
We recall some notions and results necessary for our work here.

In \cite{R2,R3,R4}, Rehren studied 2-dimensional local conformal quantum 
field theory $\B(O)$ which irreducible extends a given pair of chiral
theories $\A=\A_L\otimes \A_R$.  That is, the mathematical structure
studied there is an irreducible inclusion of nets,
$\A_L(I)\otimes \A_R(J)\subset
\B(O)$, where $I, J$ light ray intervals and $O$ a double cone
$I\times J$.  Note that here $\A_L$ and $\A_R$ can be distinct.
For such an extension, we decompose the dual canonical endomorphism
$\th$ on $\A_L\otimes \A_R$ as
$$\th=\bigoplus_{ij} Z_{ij} \a_i^L\otimes\a_j^R,$$
where $\{\a_i^L\}_i$ and $\{\a_j^R\}_j$ are systems of
irreducible DHR endomorphisms of $\A_L$ and $\A_R$,
respectively.  The matrix $Z=(Z_{ij})$ is called a coupling
matrix.  The two nets $\A_L$ and $\A_R$ define $S$- and $T$-matrices,
$S_L$, $T_L$, $S_R$, $T_R$, respectively, as in \cite{R1}.
We are interested in the case where the $S$-matrices are
invertible.  (By the results in \cite{KLM}, this invertibility,
which is called non-degeneracy of the braiding,
holds if the nets are completely rational in the
sense of \cite{KLM}.)  Then Rehren considered when the
following two intertwining relations hold.
\begin{equation}
\label{stat-symm}
T_L Z = Z T_R,\qquad\qquad S_L Z= Z S_R.
\end{equation}
Note that if $\A_L=\A_R$ and the non-degeneracy of the
braiding holds, this condition implies the usual
modular invariance of $Z$.  (We always have $Z_{00}=1$
and $Z_{ij}\in\{0,1,2,\dots\}$.)
He considered natural situations where the above equalities
(\ref{stat-symm}) hold, but also pointed out that it is not
necessarily valid in general by showing a very easy
counter-example to the intertwining property (\ref{stat-symm}).
He then continues as follows.
``A possible criterium to exclude models like the counter
examples, and hopefully to enforce the intertwining property,
could be that the local 2D theory $\B$ does not possess
nontrivial superselection sectors, but I have no proof that this
condition indeed has the desired consequences.''
M\"uger \cite{Mu} has proved that this triviality of the superselection
structures is indeed sufficient (and necessary) for
the intertwining property (\ref{stat-symm}), when the nets 
$\A_L$ and $\A_R$ are completely rational.  

\begin{theorem}[M\"uger \cite{Mu}]
\label{modular}
Under the above conditions, the following are equivalent.
\begin{enumerate}
\item The net $\B$ has only the trivial superselection sector.
\item The $\mu$-index $\mu_B$ is 1.
\item The matrix $Z$ has the intertwining property 
(\ref{stat-symm}),
$$T_L Z = Z T_R,\qquad\qquad S_L Z= Z S_R.$$
\end{enumerate}
\end{theorem}

In the case where we can naturally identify $\A_L$ and $\A_R$, 
the above theorem gives a relation between the classification
problem of the modular invariants and the classification
problem of the local extension of $\A_L\otimes \A_R$ with
$\mu$-index equal to 1.

\section{Longo-Rehren subfactors and 2-cohomology of a tensor category}

Let $M$ be a type III factor.  We say that a finite subset
$\Delta\subset\End(M)$ is a \emph{system of endomorphisms} of $M$ if
the following conditions hold, as in \cite[Definition 2.1]{BEK1}.

\begin{enumerate}
\item Each $\la\in \Delta$ is irreducible and has finite statistical
dimension.
\item The endomorphisms in $\Delta$ are mutually inequivalent.
\item We have $\id_M\in \Delta$.
\item For any $\la\in\Delta$, we have an endomorphism
$\bar\la\in\Delta$ such that $[\bar\la]$ is the conjugate sector
of $[\la]$.
\item The set $\Delta$ is closed under composition and subsequent
irreducible decomposition, i.e., for any $\la,\mu\in\Delta$,
we have non-negative integers $N_{\la,\mu}^\nu$ with
$[\la][\mu]=\sum_{\nu\in\Delta} N_{\la,\mu}^\nu [\nu]$ as sectors.
\end{enumerate}

Two typical examples of systems of endomorphism are as follows.
First, if we have a subfactor $N\subset M$ with finite index, then 
consider representatives of unitary equivalence classes of
irreducible endomorphisms appearing in irreducible decompositions
of powers $\ga^n$ of the canonical endomorphism $\ga$ for the subfactor.
If the set of representatives is finite, that is, 
if the subfactor is of finite depth, then we obtain a finite 
system of endomorphisms.  
Second, if we have a local conformal net $\A$ on the circle, we consider
representatives of unitary equivalence classes of irreducible DHR
endomorphisms of this net.  If the set of representatives is finite, 
that is, if the net is rational, then we obtain a finite
system of endomorphisms of $M=\A(I)$, where $I$ is some fixed
interval of the circle.

Recall the definition of a $Q$-system in \cite{L2}.
Let $\th$ be an endomorphism of a type III factor.
A triple $(\th, V, W)$ is called a \emph{$Q$-system} if we have
the following properties.
\begin{eqnarray}
V&\in&\Hom(\id,\th),\\
W&\in&\Hom(\th,\th^2),\\
V^* V&=&1,\\
W^* W&=&1, \label{Q0}\\
V^* W&=& \th(V^*) W\in \R_+, \label{Q1}\\
W^2&=&\th(W)W, \label{Q2}\\
\th(W^*) W &=& W W^*. \label{Q3}
\end{eqnarray}
Actually, it has been proved in \cite{LRo} that Condition
(\ref{Q3}) is redundant.  (It has been also proved in \cite{IK} that
Condition (\ref{Q2}) is redundant if (\ref{Q3}) is assumed.)
In this case, $\th$ is a canonical endomorphism of a
certain subfactor of the original factor.

For a finite system $\Delta$ as above, Longo and Rehren constructed
a subfactor $M\otimes M^\o \subset R$ 
in \cite[Proposition 4.10]{LR} such that the dual
canonical endomorphism has a decomposition
$\th=\bigoplus_{\la\in\Delta}\lambda\otimes\lambda^\op$,
by explicitly writing down a $Q$-system $(\th, V, W)$.
We, however, could have an inequivalent $Q$-system
for the same dual canonical endomorphism $\th$.
(We say that two $Q$-systems $(\th, V_1, W_1)$ and $(\th, V_2, W_2)$
are \emph{equivalent} if we have a unitary $u\in \Hom(\th,\th)$
satisfying
$$
V_2=u V_1,\qquad W_2=u\th(u)W_1 u^*.
$$
This equivalence of $Q$-systems is equivalent to inner conjugacy of
the corresponding subfactors \cite{IK}.)  We study this problem
of uniqueness of the $Q$-systems below.  Classification of $Q$-systems
for a given dual canonical endomorphism was studied as a subfactor
analogue of 2-cohomology of a group in \cite{IK}.  We show that
for a Longo-Rehren $Q$-system, we naturally have a 2-cohomology
\emph{group} of a tensor category, while 2-cohomology in \cite{IK}
is not a group in general.

Suppose we have a family $(C_{\la\mu})_{\la,\mu\in\Delta}$
with $C_{\la\mu}\in\Hom(\la\mu,\la\mu)$.  An intertwiner
$C_{\la\mu}$ naturally defines an operator
$C_{\la\mu}^\nu\in\End(\Hom(\nu,\la\mu))$ for $\nu\in\Delta$
by composition from the left.  For $\la,\mu,\nu,\pi\in\Delta$,
we have a decomposition
$$\Hom(\pi,\la\mu\nu)=\bigoplus_{\si\in\Delta}
\Hom(\si,\la\mu)\otimes\Hom(\pi,\si\nu).$$
We have
$$\bigoplus_{\si\in\Delta}C_{\la\mu}^\si\otimes C_{\si\nu}^\pi
\in\End(\Hom(\pi,\la\mu\nu))$$
according to this decomposition.
We similarly have
$$\bigoplus_{\tau\in\Delta}C_{\la\tau}^\pi\otimes C_{\mu\nu}^\tau
\in\End(\Hom(\pi,\la\mu\nu))$$
based on the last expression of the decompositions
\begin{eqnarray*}
\Hom(\pi,\la\mu\nu)&\isom&\bigoplus_{\tau\in\Delta}
\Hom(\pi,\la\tau)\otimes\la(\Hom(\tau,\mu\nu))\\
&\isom& \bigoplus_{\tau\in\Delta}
\Hom(\pi,\la\tau)\otimes\Hom(\tau,\mu\nu).
\end{eqnarray*}
We now consider the following conditions.

\begin{definition}
\label{2-cocycle}
{\rm
We say that a family $(C_{\la\mu})_{\la,\mu\in\Delta}$ 
is a \emph{unitary $2$-cocycle} of $\Delta$,
if the following conditions hold.
\begin{enumerate}
\item For $\la,\mu\in\Delta$, each $C_{\la\mu}$ is a unitary
operator in $\Hom(\la\mu,\la\mu)$.
\item For $\la\in\Delta$, we have
$C_{\la\id}=1$ and $C_{\id\la}=1$.
\label{coc1}
\item For $\la,\mu,\nu,\pi\in\Delta$, we have
$$\bigoplus_{\si\in\Delta}C_{\la\mu}^\si\otimes C_{\si\nu}^\pi=
\bigoplus_{\tau\in\Delta}C_{\la\tau}^\pi\otimes C_{\mu\nu}^\tau$$
as an identity in $\End(\Hom(\pi,\la\mu\nu))$ with respect to
the above decompositions of $\Hom(\pi,\la\mu\nu)$.
\label{coc2}
\end{enumerate}
}\end{definition}

We always assume unitarity for $C_{\la\mu}$ in this paper,
so we simply say a 2-cocycle for a unitary 2-cocycle.
For a 2-cocycle $(C_{\la\mu})_{\la,\mu\in\Delta}$,
we define $C_{\la\mu\nu}^\pi\in\End(\Hom(\pi,\la\mu\nu))$ by
$$\bigoplus_{\si\in\Delta}C_{\la\mu}^\si\otimes C_{\si\nu}^\pi.$$
Similarly, we can define
$$C_{\la_1\la_2\cdots\la_n}^{\mu_1\mu_2\cdots\mu_m}\in
\End(\Hom(\mu_1\mu_2\cdots\mu_m, \la_1\la_2\cdots\la_n)).$$ 
(Note that well-definedness
follows from the Condition 3 in Definition \ref{2-cocycle}.)
In this notation, we have $C_{\la\mu}^{\la\mu}
\in\End(\Hom(\la\mu,\la\mu))$ and this endomorphism is given as
the left multiplication of $C_{\la\mu}\in\Hom(\la\mu,\la\mu)$
on $\Hom(\la\mu,\la\mu)$, where the product structure on
$\Hom(\la\mu,\la\mu)$ is given by composition.  In this way,
we can identify $C_{\la\mu}^{\la\mu}
\in\End(\Hom(\la\mu,\la\mu))$ and
$C_{\la\mu}\in\Hom(\la\mu,\la\mu)$.

We next consider a strict $C^*$-tensor category $\T$,
with conjugates, subobjects, and direct sums, whose 
objects are given as finite direct sums of endomorphisms
in $\Delta$.  We then study an automorphism $\Phi$ of $\T$ 
such that $\Phi(\la)$ and $\la$ are unitarily equivalent for all
objects $\la$ in $\T$.   For all $\la\in\Delta$, we choose
a unitary $u_\la$ with $\Phi(\la)=\Ad(u_\la)\cdot\la$.
By adjusting $\Phi$ with $(\Ad(u_\la))_{\la\in\Delta}$, we
may and do assume that $\Phi(\la)=\la$.  Then
such an automorphism $\Phi$ gives a family of automorphisms
$$\Phi_{\la_1\la_2\cdots\la_n}^{\mu_1\mu_2\cdots\mu_m}\in
\Aut(\Hom(\mu_1\mu_2\cdots\mu_m, \la_1\la_2\cdots\la_n)),$$
for $\la_1,\la_2,\cdots,\la_n,\mu_1,\mu_2,\cdots,\mu_m\in\Delta$,
with the compatibility condition 
$$\Phi_{\la_1\la_2\cdots\la_n}^{\nu_1\nu_2\cdots\nu_k}=
\bigoplus_{\mu_1,\mu_2,\cdots,\mu_m\in\Delta}
\Phi_{\la_1\la_2\cdots\la_n}^{\mu_1\mu_2\cdots\mu_m} \otimes
\Phi_{\mu_1\mu_2\cdots\mu_m}^{\nu_1\nu_2\cdots\nu_k}$$
on the decomposition
\begin{eqnarray*}
&&\Hom(\nu_1\nu_2\cdots\nu_k, \la_1\la_2\cdots\la_n)\\
&=& \bigoplus_{\mu_1,\mu_2,\cdots,\mu_m\in\Delta}
\Hom(\mu_1\mu_2\cdots\mu_m,\la_1\la_2\cdots\la_n)\otimes
\Hom(\nu_1\nu_2\cdots\nu_k,\mu_1\mu_2\cdots\mu_m).
\end{eqnarray*}
It is clear that a family
$(C_{\la_1\la_2\cdots\la_n}^{\mu_1\mu_2\cdots\mu_m})$
arising from a 2-cocycle $(C_{\la\mu})$ is an automorphism
of a tensor category in this sense.  

Conversely, suppose
that we have an automorphism $\Phi$ of a tensor category
acting on objects trivially as above.  Then using the isomorphism
$$\Hom(\la\mu,\la\mu)\isom
\bigoplus_{\nu\in\Delta}\Hom(\nu,\la\mu)\otimes
\Hom(\la\mu,\nu),$$
the family $(\Phi_{\la\mu}^\nu)$ gives a unitary intertwiner
in $\Hom(\la\mu,\la\mu)$.  We denote this intertwiner by
$C_{\la\mu}$ and then it is clear that the family
$(C_{\la\mu})$ gives a 2-cocycle in the above sense.  Thus
in this correspondence, we can identify a 2-cocycle on $\Delta$
and an automorphism of the tensor category arising from $\Delta$
that fixes each object in the category.

We now have the following definition.

\begin{definition}
\label{vanish}
{\rm
(1) We say that 2-cocycles $(C_{\la\mu})_{\la\mu}$ and
$(C'_{\la\mu})_{\la\mu}$ are 
\emph{equivalent} if we
have a family $(\omega_\la)_\la$ of scalars of modulus 1 such that 
$$C_{\la\mu}^\nu=\omega_\nu/(\omega_\la \omega_\mu) {C'}_{\la\mu}^\nu
\in \End(\Hom(\nu,\la\mu)).$$
If a 2-cocycle $(C_{\la\mu})_{\la\mu}$ is equivalent to
$(1)_{\la\mu}$, then we say that it is \emph{trivial}.

(2) We say that a 2-cocycle $(C_{\la\mu})_{\la\mu}$ is 
\emph{scalar-valued} if all $C_{\la\mu}^\nu$'s are scalar
operators on $\Hom(\la\mu,\nu)$.

(3) We say that an automorphism $\Phi$ of the tensor category
as above is \emph{trivial} if we have a 
family $(\omega_\la)_\la$ of scalars of modulus 1 satisfying
$$\Phi_{\la_1\la_2\cdots\la_n}^{\mu_1\mu_2\cdots\mu_m}=
\om_{\mu_1}\cdots\om_{\mu_m}/(\om_{\la_1}\cdots\om_{\la_n}).$$
}\end{definition}

Note that if a $2$-cocycle is trivial, then
it is scalar-valued, in particular.

We now recall the definition of the Longo-Rehren subfactor
\cite[Proposition 4.10]{LR} as follows.
(See \cite{Ma}, \cite{O2}, \cite{P1} for related or more
general definitions.)
Let $\Delta=\{\la_k\mid k=0,1,\dots,n\}$ be a finite system
of endomorphisms of a type III factor $M$ where $\la_0=\id$.
We choose a system $\{V_k\mid k=0,1,\dots,n\}$ of isometries
with $\sum_{k=0}^n V_k V_k^*=1$ in the factor $M\otimes M^\o$,
where $M^\o$ is the opposite algebra of $M$ and we denote
the anti-linear isomorphism from $M$ onto $M^\o$ by $j$.
Then we set
$$\rho(x)=\sum_{k=0}^n V_k ((\la_k\otimes \la_k^\o) (x)) V_k^*,$$
for $x\in M\otimes M^\o$, where
$\la^\o=j\cdot\la\cdot j^{-1}$.
We set $V=V_0\in\Hom(\id,\rho)$ and define $W\in\Hom(\rho,\rho^2)$
as follows.
$$W=\sum_{k,l,m=0}^n
\sqrt{\frac{d_k d_l}{w d_m}}V_k (\la_k\otimes \la_k^\o) (V_l)
T_{kl}^m V_m^*,$$
where $d_k$ is the statistical dimension of $\la_k$,
$w$ is the global index of the system, $w=\sum_{k=0}^n d_k^2$, and
$$T_{kl}^m=\sum_{i=1}^{N_{kl}^m} (T_{kl}^m)_i\otimes 
j((T_{kl}^m)_i).$$
Here $N_{kl}^m$ is the structure constant, $\dim \Hom(\la_m,\la_k\la_l)$,
and $\{(T_{kl}^m)_i\mid i=1,2,\dots,N_{kl}^m \}$ is a fixed 
orthonormal basis of $\Hom(\la_m, \la_k\la_l)$.
Note that the operator $T_{kl}^m$ does not depend on the choice of
the orthonormal basis.  Proposition 4.10 in \cite{LR} says that
the triple $(\rho, V, W)$ is a $Q$-system.
Thus we have a subfactor $M\otimes M^\o\subset R$ with index $w$
corresponding to
the dual canonical endomorphism $\rho$.  We call this a
\emph{Longo-Rehren subfactor} arising from the system $\Delta$.

Furthermore, if $\Delta$ is a subsystem of all the irreducible
DHR endomorphisms of a local conformal net $\A$, then any $Q$-system
having this dual canonical endomorphism gives an extension
$\B\supset \A\otimes\A^\op$.  This 2-dimensional net $\B$ is
local if and only if $\e(\rho,\rho)W=W$
by \cite[Proposition 4.10]{LR}, where $\e$ is the braiding.
In general, if the system $\Delta$ has a braiding $\e$, and
this condition $\e(\rho,\rho)W=W$ holds, we say that
the $Q$-system $(\rho,V,W)$ satisfies \emph{locality}.

We now would like to characterize a general $Q$-system having the
same dual canonical endomorphism $\rho$.  First, we have the following
simple lemma.

\begin{lemma}\label{oper}
Let $F, F'$ be finite dimensional complex Hilbert spaces and
$j$ an anti-linear isomorphism from $F$ onto $F'$.  
For any vector $\xi\in F\otimes F'$, we define
a linear map $A: F \to F$ by $\xi=\sum_k A\xi_k\otimes j(\xi_k)$
where $\{\xi_k \}$ is an orthonormal basis of $F$.  Then
this linear map $A$ is independent of the choice of the
orthonormal basis $\{\xi_k \}$.
\end{lemma}

\begin{proof}
This is straightforward by the anti-isomorphism property of $j$.
\end{proof}

The next Theorem gives our characterization of $Q$-systems.

\begin{theorem}
\label{Q-sys}
Let $\Delta, \rho, V, W$ be as above.  If another triple
$(\rho, V, W')$ with $W'\in\Hom(\rho,\rho^2)$ is a $Q$-system, 
we have a $2$-cocycle $(C_{\la\mu})_{\la,\mu\in\Delta}$ such that
\begin{equation}
W'=\sum_{k,l,m=0}^n
\sqrt{\frac{d_k d_l}{w d_m}} V_k (\la_k\otimes \la_k^\o) (V_l)
(C_{\la_k \la_l}\otimes 1)T_{kl}^m V_m^*.
\label{Q'-sys}
\end{equation}

Conversely, if we have a $2$-cocycle $(C_{\la\mu})_{\la,\mu\in\Delta}$,
then the triple $(\rho, V, W')$ with $W'$ defined as in
(\ref{Q'-sys}) is a $Q$-system.

The $Q$-system $(\rho, V, W')$ is equivalent
to the above canonical
$Q$-system $(\rho, V, W)$ if and only if the
corresponding  $2$-cocycle
$(C_{\la\mu})_{\la,\mu\in\Delta}$ is trivial, if and only
if the corresponding automorphism of the tensor category
arising from $\Delta$ is trivial.

Moreover, suppose that the system $\Delta$ has a braiding
$\e^\pm$.  Then the $Q$-system $(\rho, V, W')$ satisfies locality
if and only if the corresponding $2$-cocycle
$(C_{\la\mu})_{\la,\mu\in\Delta}$ satisfies the following
symmetric condition.
\begin{equation}\label{symmetric}
C_{\la\mu}=\e^-_{\mu\la}C_{\mu\la}\e^+_{\la\mu}, 
\end{equation}
for all $\la,\mu\in\Delta$.  If this symmetric condition
holds, the corresponding automorphism of the tensor
category arising from $\Delta$ is an automorphism of a 
braided tensor category.
\end{theorem}

\begin{proof}
If $(\rho, V, W')$ with $W'\in\Hom(\rho,\rho^2)$ is a $Q$-system,
then we have a system of intertwiners
$(C_{\la\mu})_{\la,\mu\in\Delta}$ such that identity (\ref{Q'-sys})
holds and the intertwiners $(C_{\la\mu})$ are uniquely
determined by Lemma \ref{oper}.  
Expanding the both hand sides of identity (\ref{Q2}), we obtain
the following identity.
\begin{eqnarray}\label{expand}
&&\sum_{k,l,m,p,q=0}^n
\sqrt{\frac{d_k d_l d_m}{w^2 d_p}}
V_k (\la_k\otimes \la_k^\o) (V_l)
(\la_k\la_l\otimes \la_k^\o\la_l^\o) (V_m) \nonumber \\
&&\quad (\la_k\otimes \la_k^\o)((C_{\la_l \la_m}\otimes 1)T_{lm}^q)
(C_{\la_k \la_q}\otimes 1)T_{kq}^p V_p^* \nonumber \\
&=& \sum_{k,l,m,p,r=0}^n
\sqrt{\frac{d_k d_l d_m}{w^2 d_p}}
V_k (\la_k\otimes \la_k^\o) (V_l)
(\la_k\la_l\otimes \la_k^\o\la_l^\o) (V_m) \nonumber \\
\quad &&(C_{\la_k \la_l}\otimes 1)T_{kl}^r
(C_{\la_r \la_m}\otimes 1)T_{rm}^p V_p^*.
\end{eqnarray}
We decompose
\begin{eqnarray*}
\Hom(\la_p,\la_k \la_l \la_m)&\isom&
\bigoplus_{q=0}^n \Hom(\la_p,\la_k \la_q)\otimes 
\Hom(\la_q,\la_l \la_m)\\
&\isom& \bigoplus_{r=0}^n \Hom(\la_r,\la_k \la_l)\otimes 
\Hom(\la_p,\la_r \la_m),
\end{eqnarray*}
as above, and apply Lemma \ref{oper}
to the above identity (\ref{expand})
to obtain Condition \ref{coc2} in Definition \ref{2-cocycle}.
Similarly, Condition \ref{coc1} in Definition \ref{2-cocycle} 
follows from identity (\ref{Q1}).  

We next prove unitarity of $C_{\la\mu}\in\Hom(\la\mu,\la\mu)$.
First note that the operator $C_{\la\bar\la}^\id, C_{\bar\la\la}^\id$ 
are scalar multiples of the identity because $\Hom(\id,\la\bar\la)$,
$\Hom(\id,\bar\la\la)$ are both 1-dimensional.

Since the triple $(\rho, V, W')$ also satisfies identity (\ref{Q3}),
we expand the both side hands of identity (\ref{Q3}) and
use Lemma \ref{oper} as in the above arguments.
Then we obtain the following.
The intertwiner space $\Hom(\la\mu,\nu\si)$ for 
$\la,\mu,\nu,\si\in\Delta$ can be decomposed in two ways as
follows.
\begin{eqnarray}
\Hom(\la\mu,\nu\si) &\isom &
\bigoplus_{\tau\in\Delta} \Hom(\la,\nu\tau)\otimes\Hom(\tau\mu,\si)\\
&\isom& 
\bigoplus_{\pi\in\Delta} \Hom(\la\mu,\pi)\otimes\Hom(\pi,\nu\si).
\label{decompo2}
\end{eqnarray}
On one hand, Lemma \ref{oper} applied to the left hand side of 
identity (\ref{Q3}) produces a map in
$\End(\Hom(\la\mu,\nu\si))$ which maps 
$T_i\otimes S_j^*\in \Hom(\la,\nu\tau)\otimes\Hom(\tau\mu,\si)$,
identified with $\nu(S_j^*)T_i\in \Hom(\la\mu,\nu\si)$, to
$\nu(S_j^* C_{\tau\mu}^*) C_{\nu\tau} T_i\in \Hom(\la\mu,\nu\si)$,
where $T_i$ and $S_j$ are isometries in
$\Hom(\la,\nu\tau)$ and $\Hom(\si,\tau\mu)$, respectively.
On the other hand, Lemma \ref{oper} applied to the right hand side of 
identity (\ref{Q3}) produces a map in
$\End(\Hom(\la\mu,\nu\si))$ which maps 
${T'_i}^*\otimes S'_j\in \Hom(\la\mu,\pi)\otimes\Hom(\pi,\nu\si)$,
identified with $S'_j {T'_i}^*\in \Hom(\la\mu,\nu\si)$, to
$C_{\nu\si} S'_j {T'_i}^* C_{\la\mu}^*\in \Hom(\la\mu,\nu\si)$,
where $T'_i$ and $S'_j$ are isometries in
$\Hom(\pi,\la\mu)$ and $\Hom(\pi,\nu\si)$, respectively.
These two maps are equal in $\End(\Hom(\la\mu,\nu\si))$.
In the above decomposition (\ref{decompo2}), we set
$\la=\si=\id$ and $\mu=\nu$, then we have $\tau=\bar\mu$
and $\pi=\mu$ in the summations.  
With Frobenius reciprocity as in \cite{I2} 
and  the above identity of two maps in $\End(\Hom(\la\mu,\nu\si))$,
we obtain the identity
\begin{equation}
C_{\mu\bar\mu}^\id \overline{C_{\bar\mu\mu}^\id} =1.
\label{modulus}
\end{equation}

We next apply identity (\ref{Q0}) to (\ref{Q'-sys}) and obtain
the following equality
\begin{equation}
\sum_{\la,\mu\in\Delta} d_\la d_\mu K_{\la\mu}^\nu = w d_\nu,
\label{dim-id}
\end{equation}
where we have set 
$K_{\la\mu}^\nu=\Tr((C_{\la\mu}^\nu)^* C_{\la\mu}^\nu)$ 
and $\Tr$ is the
non-normalized trace  on $\Hom(\la\mu,\la\mu)$.
Setting $\nu=\id$ in (\ref{dim-id}), we obtain
$$\sum_{\la\in\Delta} d_\la^2 |C_{\la\bar\la}^\id|^2 = w,$$
which, together with (\ref{modulus}),
implies $|C_{\la\bar\la}^\id|=1$ for all $\la\in\Delta$.

In the above decomposition (\ref{decompo2}), we now set
$\la=\id$, then we have $\tau=\bar\nu$
and $\pi=\mu$ in the summations.
With Frobenius reciprocity as in \cite{I2}
and  the above identity of two maps in $\End(\Hom(\la\mu,\nu\si))$,
we obtain the identity
\begin{equation}
C_{\nu\bar\nu}^\id \nu((C_{\bar\nu\mu}^\si)^* \tilde T)
R_{\nu\bar\nu} = \sqrt{\frac{d_\mu}{d_\nu d_\si}} C_{\nu\si}^\mu T,
\label{WW*}
\end{equation}
for all $T\in \Hom(\mu,\nu\si)$, where $\tilde T\in
\Hom(\bar\nu\mu,\si)$ is the Frobenius dual of $T$
and $R_{\nu\bar\nu}\in\Hom(\id,\nu\bar\nu)$ is
the canonical isometry.
This identity (\ref{WW*}), Condition \ref{coc2} in 
Definition \ref{2-cocycle}, already
proved, and identity (\ref{modulus}) imply the following identity,
\begin{eqnarray*}
\lan C_{\bar\nu\mu}^\si T, C_{\bar\nu\mu}^\si S \ran &=&
(C_{\nu\bar\nu}^\id)^* R_{\bar\nu \nu}^*
\bar\nu(C_{\nu\si}^\mu \tilde S^*) C_{\bar\nu\mu}^\si T\\
&=& (C_{\nu\bar\nu}^\id)^* C_{\bar\nu\nu}^\id S^* T\\
&=& \lan T, S\ran,
\end{eqnarray*}
where we have $T,S\in \Hom(\si,\bar\nu\mu)$ and the inner product
is given by $\lan T, S\ran=S^* T\in\C$.
This is the desired unitarity of $C_{\bar\nu\mu}$.

The converse also holds in the same way and the remaining parts
are straightforward.
\end{proof}

It is easy to see that we can multiply 2-cocycles and
the multiplication on the equivalences classes of 2-cocycles
is well-defined.  In this way, we obtain a group and this
is called the \emph{$2$-cohomology group} of $\Delta$ (or of
the corresponding tensor category).  It is also easy to 
see that the multiplication gives the composition
of the corresponding automorphisms of the tensor category.

The part of the above theorem on a bijective
correspondence between $Q$-systems $(\rho,V,W')$ with
locality and automorphisms of the braided tensor category
has been also announced by M\"uger in \cite{Mu}.

\begin{example}\label{group-case}
{\rm
If all the endomorphisms in $\Delta$ are automorphisms, then
the fusion rules determine a finite group $G$.  It is easy
to see that the Longo-Rehren $Q$-system gives a crossed
product by an outer action of $G$ and the above 2-cohomology group
for $\Delta$ is isomorphic to the usual 2-cohomology group
of $G$.

Furthermore, if the system $\Delta$ has a braiding, then the group
$G$ is abelian.  In this case, the symmetric condition of a 
cocycle means $c_{g,h}=c_{h,g}$ for the corresponding
usual 2-cocycle $c$ of 
the finite abelian group $G$.  It is well-known that such a 
2-cocycle is trivial.  (See \cite[Lemma 3.4.2]{B}, for example.)
}\end{example}

When all the 2-cocycles for $\Delta$ are trivial,
we say that we have a 2-cohomology vanishing for $\Delta$.
Thus, 2-cohomology vanishing implies uniqueness of the Longo-Rehren
subfactor in the following sense.

\begin{corollary}\label{coro}
Let $\Delta$ be as above.  If we have a 2-cohomology vanishing 
for $\Delta$ and 
$\rho=\bigoplus_{\la\in\Delta} \la\otimes \la^\op$ is a dual
canonical endomorphism for a subfactor $M\otimes M^\op\subset P$, 
then this subfactor is inner conjugate to the Longo-Rehren subfactor
$M\otimes M^\op\subset R$.
\end{corollary}

\section{2-cohomology vanishing and classification}

In this section, we first study a general theory of 2-cohomology
for a $C^*$-tensor category and then apply it to the tensor
categories related to the Virasoro algebra.
We consider a strict $C^*$-tensor category $\T$
(with conjugates, subobjects, and direct sums) in the
sense of \cite{DR,LRo} and we assume that we have only
finitely many equivalence classes of irreducible objects in
$\T$ and that each object has a decomposition into a finite
direct sum of irreducible objects.
Such a tensor category is often called \emph{rational}.
We may and do assume that our tensor
categories are realized as
those of  endomorphisms of a type III factor.
Choose a system $\Delta$ of endomorphisms of a type III
factor $M$ corresponding to the $C^*$-tensor category $\T$.
Suppose we have a 2-cocycle $(C_{\la\mu})_{\la,\mu\in\Delta}$.

We introduce some basic notions.
Suppose that we have $\si\in\Delta$ such that for any $\la\in\Delta$,
there exists $k\ge 0$ such that $\la \prec \si^k$.  Then we say
that $\si$ is a \emph{generator} of $\Delta$.  In the following,
we consider only the case $\si=\bar\si$.  In this
case, we say that \emph{$\Delta$ has a self-conjugate generator $\si$.}

Suppose $\si\in\Delta$ is a self-conjugate
generator of $\Delta$.  We further
assume that for all $\la, \mu\in\Delta$, we have
$\dim \Hom(\la\si,\mu)\in\{0,1\}$.
In this case, we say that \emph{multiplications by $\si$ have
no multiplicities}.

Take $\la_1,\la_2,\la_3,\la_4 \in \Delta$ and assume
\begin{eqnarray*}
&&\dim \Hom(\la_1\si,\la_2)\;=\;\dim \Hom(\si\la_1,\la_3)\\
&=&\dim \Hom(\la_3\si,\la_4)\;=\;\dim \Hom(\si\la_2,\la_4)\;=\;1.
\end{eqnarray*}
Choose isometric intertwiners
\begin{eqnarray*}
&&T_1\in\Hom(\la_2,\la_1\si),\quad T_2\in\Hom(\la_4,\si\la_2),\\
&&T_3\in\Hom(\la_3,\si\la_1),\quad T_4\in\Hom(\la_4,\la_3\si).
\end{eqnarray*}
Then the composition
$T_4^* T_3^* \si(T_1) T_2$ is in $\Hom(\la_4,\la_4)=\C$.
This values is the \emph{connection} as in \cite{O1},
\cite[Chapter 9]{EK}.
We denote this complex number by $W(\la_1,\la_2,\la_3,\la_4)$.
(Note that this value depends on $T_1,T_2,T_3,T_4$ though they
do not appear in the notation.)
If all these complex numbers are non-zero, then
we say that \emph{the connections of $\Delta$ with respect to the
generator $\si$ are non-zero.}  This condition is
independent of the choices of isometric intertwiners $T_j$'s,
because we now assume that multiplications by $\si$ have
no multiplicities.

Suppose we have a map $g:\Delta \to \Z/2\Z$.  For
an endomorphism $\si$ that is a direct sum of elements
$\la_j$'s with $g(\la_j)=k\in\Z/2\Z$, we also set
$g(\si)=k$.  If we have 
$g(\la\mu)=g(\la)+g(\mu)$, then we say that $\Delta$ has a
$\Z/2\Z$-grading.  An endomorphism $\la\in\Delta$ is
called \emph{even} [resp. \emph{odd}] when $g(\la)=0$
[resp. $g(\la)=1$].

\begin{theorem}
\label{vanishing}
Suppose we have a finite system $\Delta$ of endomorphisms with
a self-conjugate generator $\si\in\Delta$ 
satisfying all the following conditions.
\begin{enumerate}
\item Multiplications by $\si$ have no multiplicities.
\item One of the following holds.
\begin{enumerate}
\item We have $\si\prec\si^2$.
\item The system $\Delta$ has a
$\Z/2\Z$-grading and the generator $\si$ is odd.
\end{enumerate}
\item The connections of $\Delta$ with respect to the
generator $\si$ are non-zero.
\item For any $\la,\nu_1,\nu_2\in\Delta$ with
$\nu_1\prec\si^n$, $\nu_2\prec\si^n$, $\la\prec\si\nu_1$, and
$\la\prec\si\nu_2$, we have $\mu\in\Delta$ with
$\mu\prec\si^{n-1}$, $\nu_1\prec\si\mu$, and $\nu_2\prec\si\mu$.
\end{enumerate}
Then any 2-cocycle $(C_{\la\mu})_{\la\mu}$ of $\Delta$ is trivial.
\end{theorem}

Before presenting a proof, we make a comment on Condition 4.
Consider the Bratteli diagram for the higher relative commutants
of a subfactor $\si(M)\subset M$.  We number the steps of the
Bratteli diagrams as $0,1,2,\dots$.  Then Condition 4 says the following.
(Recall that $\si$ is self-conjugate.)
Suppose we have vertices corresponding to $\nu_1$ and $\nu_2$ at the
$n$-th step of the Bratteli diagrams, and they are connected to the
vertex $\la$ in the $n+1$-st step.  Then there exists a vertex $\mu$
in the $n-1$-st step that is connected to $\nu_1$ and $\nu_2$.
Note that if $\nu_1$ and $\nu_2$ already appear in the $n-2$-nd step,
then this condition trivially holds by taking $\mu=\la$.  Thus,
if the subfactor $\si(M)\subset M$ is of finite depth, then checking
finitely many cases is sufficient for verifying Condition 4,
and this can be done by drawing the principal graph of the subfactor
$\si(M)\subset M$.

\begin{proof}
Using Conditions 1, 3 and 4, we first prove that the unitary operator
$$C_{\si\si\cdots\si}^\la\in\End(\Hom(\la,\si\si\cdots\si))$$
is scalar for any $\la\in\Delta$.
Let the number of $\si$'s in $C_{\si\si\cdots\si}^\la$ be $k$ and
we prove the above property $C_{\si\si\cdots\si}^\la\in\C$ by
induction on $k$.  Note that the intertwiner space
$\Hom(\la,\si\si\cdots\si)$ is decomposed as
$$\bigoplus \Hom(\la_1,\si\si)\otimes\Hom(\la_2,\la_1\si)\otimes
\cdots\otimes \Hom(\la,\la_{k-2}\si),$$
and each of the space
$$\Hom(\la_1,\si\si)\otimes\Hom(\la_2,\la_1\si)\otimes
\cdots\otimes \Hom(\la,\la_{k-2}\si)$$
is one-dimensional by Condition 1.  Each such one-dimensional
subspace gives a non-zero eigenvector
of the unitary operator $C_{\si\si\cdots\si}^\la$ with eigenvalue
$$C_{\si\si}^{\la_1}C_{\la_1\si}^{\la_2}\cdots C_{\la_{k-2}\si}^{\la}$$
and what we have to
prove is these eigenvalues are all identical.
Note that the decomposition of $\Hom(\la,\si\si\cdots\si)$ as above
is depicted graphically in Figure \ref{decomp1}.  Another picture
Figure \ref{decomp2} gives another decomposition into a direct sum
of one-dimensional eigenspaces.  Roughly speaking, what we prove
is that if a unitary matrix has several ``different'' decompositions
into direct sums of one-dimensional eigenspaces, then the unitary
matrix need to be a scalar multiple of the identity matrix.

\begin{figure}[htb]
\begin{center}
\unitlength 0.6mm
\begin{picture}(160,100)
\thicklines 
\put(10,10){\circle*{1}}
\put(30,10){\circle*{1}}
\put(50,10){\circle*{1}}
\put(70,10){\circle*{1}}
\put(150,10){\circle*{1}}
\put(20,20){\circle*{1}}
\put(30,30){\circle*{1}}
\put(40,40){\circle*{1}}
\put(50,50){\circle*{1}}
\put(70,70){\circle*{1}}
\put(80,80){\circle*{1}}
\put(80,90){\circle*{1}}
\put(10,10){\line(1,1){40}}
\put(70,70){\line(1,1){10}}
\put(20,20){\line(1,-1){10}}
\put(30,30){\line(1,-1){20}}
\put(40,40){\line(1,-1){30}}
\put(80,80){\line(1,-1){70}}
\put(80,80){\line(0,1){10}}
\put(110,10){\makebox(0,0){$\cdots$}}
\put(60,60){\makebox(0,0){$\cdots$}}
\put(10,6){\makebox(0,0){$\sigma$}}
\put(30,6){\makebox(0,0){$\sigma$}}
\put(50,6){\makebox(0,0){$\sigma$}}
\put(70,6){\makebox(0,0){$\sigma$}}
\put(150,6){\makebox(0,0){$\sigma$}}
\put(21,28){\makebox(0,0){$\la_1$}}
\put(31,38){\makebox(0,0){$\la_2$}}
\put(41,48){\makebox(0,0){$\la_3$}}
\put(68,78){\makebox(0,0){$\la_{k-2}$}}
\put(80,95){\makebox(0,0){$\la$}}
\end{picture}
\end{center}
\caption{Decomposition into a direct sum of one-dimensional eigenspaces}
\label{decomp1}
\end{figure}

\begin{figure}[htb]
\begin{center}
\unitlength 0.6mm
\begin{picture}(160,100)
\thicklines 
\put(150,10){\circle*{1}}
\put(130,10){\circle*{1}}
\put(110,10){\circle*{1}}
\put(90,10){\circle*{1}}
\put(10,10){\circle*{1}}
\put(140,20){\circle*{1}}
\put(130,30){\circle*{1}}
\put(120,40){\circle*{1}}
\put(110,50){\circle*{1}}
\put(90,70){\circle*{1}}
\put(80,80){\circle*{1}}
\put(80,90){\circle*{1}}
\put(150,10){\line(-1,1){40}}
\put(90,70){\line(-1,1){10}}
\put(140,20){\line(-1,-1){10}}
\put(130,30){\line(-1,-1){20}}
\put(120,40){\line(-1,-1){30}}
\put(80,80){\line(-1,-1){70}}
\put(80,80){\line(0,1){10}}
\put(50,10){\makebox(0,0){$\cdots$}}
\put(100,60){\makebox(0,0){$\cdots$}}
\put(150,6){\makebox(0,0){$\sigma$}}
\put(130,6){\makebox(0,0){$\sigma$}}
\put(110,6){\makebox(0,0){$\sigma$}}
\put(90,6){\makebox(0,0){$\sigma$}}
\put(10,6){\makebox(0,0){$\sigma$}}
\put(139,28){\makebox(0,0){$\la_1$}}
\put(129,38){\makebox(0,0){$\la_2$}}
\put(119,48){\makebox(0,0){$\la_3$}}
\put(92,78){\makebox(0,0){$\la_{k-2}$}}
\put(80,95){\makebox(0,0){$\la$}}
\end{picture}
\end{center}
\caption{Decomposition into a direct sum of one-dimensional eigenspaces}
\label{decomp2}
\end{figure}

First, let $k=2$.  By Condition 1, the space $\Hom(\la,\si\si)$ is
one-dimensional for any $\la\in\Delta$, so we obviously have
$C_{\si\si}^\la\in\C$.

Suppose now we have $C_{\si\si\cdots\si}^\la\in\C$ for any
$\la\in\Delta$ if the number of $\si$'s is less than or equal to
$k$.  We will prove
$C_{\si\si\cdots\si}^\la\in\C$ for any
$\la\in\Delta$ when the number of $\si$'s is $k+1$.  First note
that we have $C_{\si\si\cdots\si}^\la C_{\la\si}^\mu\in\C$
by the induction hypothesis and Condition 1.  What we have to prove
is that this scalar is independent of $\la$ when $\mu$ is fixed.
That is, suppose we have $\la,\la',\mu\in\Delta$, $\la\prec\si^k$,
$\la'\prec\si^k$, $\mu\prec\la\si$, $\mu\prec\la'\si$.
We will prove
$$C_{\si\si\cdots\si}^\la C_{\la\si}^\mu=
C_{\si\si\cdots\si}^{\la'} C_{\la'\si}^\mu\in\C.$$
By Condition 4, there exists $\nu\in\Delta$ such that
$\nu\prec\si^{k-1}$, $\la\prec\si\nu$, and $\la'\prec\si\nu$.
Then there exists $\tau\in\Delta$ such that
$\tau\prec\nu\si$ and $\mu\prec\si\tau$.
Note that we have
$$C_{\si\si\cdots\si}^\la C_{\la\si}^\mu=
C_{\si\nu}^\la C_{\si\si\cdots\si}^\nu C_{\la\si}^\mu\in\C,$$
where the number of $\si$'s in $C_{\si\si\cdots\si}^\nu$ is $k-1$.
The scalar $C_{\si\nu}^\la C_{\la\si}^\mu$ is the eigenvalue
of the operator $C_{\si\nu\si}^\mu$ corresponding to the
eigenvector given by the one-dimensional intertwiner space
$\Hom(\la,\si\nu)\otimes\Hom(\mu,\la\si)$.  Similarly,
the scalar $C_{\nu\si}^\tau C_{\si\tau}^\mu$ is the eigenvalue
of the same operator $C_{\si\nu\si}^\mu$ corresponding to the
eigenvector given by the one-dimensional intertwiner space
$\Hom(\tau,\nu\si)\otimes\Hom(\mu,\si\tau)$.  Condition 3
implies that these two eigenvectors are not orthogonal, thus
the two eigenvalues are equal, because the operator 
$C_{\si\nu\si}^\mu$ has an orthonormal basis of eigenvectors and
thus it is normal.  In this way, we obtain the
identities
$$C_{\si\nu}^\la C_{\la\si}^\mu =
C_{\nu\si}^\tau C_{\si\tau}^\mu=
C_{\si\nu}^{\la'} C_{\la'\si}^\mu,$$
which implies
$$C_{\si\si\cdots\si}^\la C_{\la\si}^\mu=
C_{\si\nu}^\la C_{\si\si\cdots\si}^\nu C_{\la\si}^\mu=
C_{\si\nu}^{\la'} C_{\si\si\cdots\si}^\nu C_{\la'\si}^\mu=
C_{\si\si\cdots\si}^{\la'} C_{\la'\si}^\mu\in\C,$$
as desired, where the numbers of $\si$'s in 
$C_{\si\si\cdots\si}^\la$, $C_{\si\si\cdots\si}^\nu$, and
$C_{\si\si\cdots\si}^{\la'}$ are $k$, $k-1$, and $k$,
respectively.

We next prove the triviality of the cocycle $C$ by using
Condition 2.  

First we assume we have 2 (a) of the assumptions in the Theorem,
that is, $\si\prec\si^2$.  Set $\om_\id=1$.
Since $\id\prec\si^2$, the condition
$C_{\si\si\si}^\si\in\C$ implies that
$C_{\si\si}^\si C_{\si\si}^\si = C_{\si\si}^\id C_{\id\si}^\si$.
By unitariry of $C$ in Theorem \ref{Q-sys}, we have
$|C^\si_{\si\si}|=1$, 
we thus set $\om_\si=(C^\si_{\si\si})^{-1}\in\C$.
(Recall that we have already proved $C^\si_{\si\si}$ is a scalar.)
Then this implies $\C_{\si\si}^\id=\om_\id/\om_\si^2$.  For 
$\la\in \Delta$ not equivalent to $\id,\si$, we choose a minimum
positive integer $k$ with $\la\prec\si^k$.  We set
$\om_\la=\om_\si^k C_{\si\cdots\si}^\la\in\C$, where the number of
$\si$'s in $C_{\si\cdots\si}^\la$ is $k$.  For any $m > k$, we
can represent the scalar $C_{\si\cdots\si}^\la$, where $\si$ appears
for $m$ times, as
$C_{\si\si}^\si \cdots C_{\si\si}^\si C_{\si\cdots\si}^\la$, where
the number of $C_{\si\si}^\si$'s is $m-k$ and the number of
$\si$'s in $C_{\si\cdots\si}^\la$ is $k$.  This implies
$C_{\si\cdots\si}^\la = \om_\la/\om_\si^m$, where the number of
$\si$'s in $C_{\si\cdots\si}^\la$ is $m$.   Now choose arbitrary
$\la,\mu,\nu\in\Delta$ with $\la\prec\si^l$, $\mu\prec\si^m$.
We can represent $C_{\si\cdots\si}^\nu\in\C$ with $\si$ appearing for
$l+m$ times, as the product
$C_{\la\mu}^\nu C_{\si\cdots\si}^\la C_{\si\cdots\si}^\mu$ where
$\si$'s appear for $l$ and $m$ times, respectively, and then we obtain
$$C_{\la\mu}^\nu \frac{\om_\la}{\om_\si^l}\frac{\om_\mu}{\om_\si^m}
=\frac{\om_\nu}{\om_\si^{l+m}},$$
which gives $C_{\la\mu}^\nu \om_\la \om_\mu = \om_\nu$.  
Unitarity in Theorem \ref{Q-sys} gives $\om_\la \om_\mu\neq0$,
we thus have $C_{\la\mu}^\nu=\om_\nu/(\om_\la \om_\mu)$.

We next deal with the the case 2 (b), that is, we now assume that
the system $\Delta$ has a
$\Z/2\Z$-grading and the generator $\si$ is odd.  We first set
$\om_\id=1$.  Since
$\id\prec\si^2$, we next set $\om_\si$ to be a square root of
$(C_{\si\si}^\id)^{-1}$.  (Note that $|C_{\si\si}^\id|=1$
by unitarity in Theorem \ref{Q-sys}.)
It does not matter which square root we choose.  For 
$\la\in \Delta$ not equivalent to $\id,\si$, we choose a minimum
positive integer $k$ with $\la\prec\si^k$ in the same way as above
in the case of 2 (a).  We again set
$\om_\la=\om_\si^k C_{\si\cdots\si}^\la\in\C$, where the number of
$\si$'s in $C_{\si\cdots\si}^\la$ is $k$.  For any $m > k$, we
can represent the scalar $C_{\si\cdots\si}^\la$, where $\si$ appears
for $m$ times, as
$C_{\si\si}^\id \cdots C_{\si\si}^\id C_{\si\cdots\si}^\la$, where
the number of $C_{\si\si}^\id$'s is $(m-k)/2$ and the number of
$\si$'s in $C_{\si\cdots\si}^\la$ is $k$, because $m-k$ is now even,
due to the $\Z/2\Z$-grading.  Then we obtain
$$C_{\si\cdots\si}^\la = \frac{1}{\om_\si^{m-k}}
\frac{\om_\la}{\om_\si^k} =\frac{\om_\la}{\om_\si^m},$$
where the number of $\si$'s in $C_{\si\cdots\si}^\la$ is $m$.
Then the same argument as in the above case of 2 (a) proves
the triviality of the cocycle $C_{\la\mu}$.
\end{proof}

\begin{remark}{\rm
The 2-cohomology does not vanish in general, as well-known in
the finite group case.  For example, if the system
$\Delta$ arises from an outer action of a finite group 
$G=\Z/2\Z\times \Z/2\Z$, it is known that we have a
non-trivial unitary 2-cocycle for this group $G$.
So as in Example \ref{group-case},
the 2-cohomology for the corresponding tensor category
does not vanish.
}\end{remark}

In \cite[Theorem 2.4, Theorem 4.1]{KL}, we have
classified local extensions of the conformal nets $SU(2)_k$ and
$\Vir_c$ with $k=1,2,3,\dots$ and $c=1-6/m(m+1)$, $m=2,3,4,\dots$.
(Here the symbol $\Vir_c$ denotes the Virasoro net with central
charge $c$.)  We use the symbols $SU(2)_k$ and $\Vir_c$ also
for the corresponding $C^*$-tensor categories.  We also say that
the corresponding $C^*$-tensor categories of these local extensions
of the nets $SU(2)_k$ and $\Vir_c$ are \emph{extensions} of the
tensor categories $SU(2)_k$ and $\Vir_c$.  Furthermore, the 
tensor category $SU(2)_k$ has a natural $\Z/2\Z$-grading and 
the even objects make a sub-tensor category.  
We call it \emph{the even part of $SU(2)_k$}.
We then have the following theorem.

\begin{theorem}
\label{SU2-Virasoro}
Any finite system $\Delta$ of endomorphisms corresponding to
one of the following tensor categories has a self-conjugate
generator $\si$
satisfying all the Conditions in Theorem \ref{vanishing}, and
thus, we have 2-cohomology vanishing for these tensor categories.
\begin{enumerate}
\item The $SU(2)_k$-tensor categories and their extensions.
\item The sub-tensor categories of those in Case 1.
\item The Virasoro tensor
categories $\Vir_c$ with $c<1$ and their extensions.
\item The sub-tensor categories of those in Case 3.
\end{enumerate}
\end{theorem}

\begin{proof}
We deal with the following cases separately.  Here for the extensions
of $SU(2)_k$-tensor categories and the Virasoro tensor categories
$\Vir_c$, we use the labels by (pairs of) Dynkin diagrams as in 
\cite[Theorem 2.4, Theorem 4.1]{KL}, which arise from the labels
of modular invariants by Cappelli-Itzykson-Zuber \cite{CIZ}.
(These also correspond to the type I modular invariants
listed in Table \ref{CIZ-mod} in this paper.)
Note that the braiding
does not matter now, so we ignore the braiding structure here.
\begin{enumerate}
\item The $SU(2)_k$-tensor categories and their extensions.
\begin{enumerate}
\item Tensor categories $A_n$.
\item Tensor categories $D_{2n}$.
\item Tensor category $E_6$.
\item Tensor category $E_8$.
\end{enumerate}
\item The (non-trivial) sub-tensor categories of those in Case 1.
\begin{enumerate}
\item The group $\Z/2\Z$.
\item The even parts of the $SU(2)_k$-tensor categories.
\end{enumerate}
\item The Virasoro tensor categories $\Vir_c$ with $c<1$
and their extensions.
\begin{enumerate}
\item Tensor categories $(A_{n-1}, A_n)$.
\item Tensor categories $(A_{4n}, D_{2n+2})$.
\item Tensor categories $(D_{2n+2}, A_{4n+2})$.
\item Tensor category $(A_{10},E_6)$.
\item Tensor category $(E_6, A_{12})$.
\item Tensor category $(A_{28},E_8)$.
\item Tensor ategory $(E_8,A_{30})$.
\end{enumerate}
\item The (non-trivial) sub-tensor categories of those in Case 3.
\begin{enumerate}
\item The sub-tensor categories of those in Case 3 (a).
\item The sub-tensor categories of those in Case 3 (b).
\item The sub-tensor categories of those in Case 3 (c).
\item The sub-tensor categories of those in Case 3 (d).
\item The sub-tensor categories of those in Case 3 (e).
\item The sub-tensor categories of those in Case 3 (f).
\item The sub-tensor categories of those in Case 3 (g).
\end{enumerate}
\end{enumerate}

Case 1 (a).  We label the irreducible objects of the tensor category
$A_{k+1}$ with $0,1,2,\dots, k$, as usual.  Let $\si$ be the
standard generator $1$.  Condition 1 of Theorem \ref{vanishing}
clearly holds.  Since the fusion rule of the tensor
category $SU(2)_k$ has
a $\Z/2\Z$-grading and this generator $1$ is odd, Condition 2 (b)
also holds.  Now the connection values with respect to this $\si$
are the usual connection values of the paragroup $A_{k+1}$ as in
\cite{O1}, \cite{K1}, \cite[Section 11.5]{EK}, and they are non-zero
and Condition 3 holds.
The multiplication rule by the generator $\si$
is described with the usual Bratteli diagram for the principal
graph $A_{k+1}$ as in \cite{J}, \cite[Chapter 9]{EK}, so we see
that Condition 4 holds.

Case 1 (b).  The irreducible objects of the tensor category
are labeled with the even vertices of the Dynkin diagram $D_{2n}$.
(So we also use the name $D_{2n}^{\rm even}$ for this tensor category.)
If $2n=4$, then this tensor category is given by the group $\Z/3\Z$, and
we can verify the conclusion directly, so we assume that $n > 2$.
We label $\si$ as in Figure \ref{D2n}.

\begin{figure}[htb]
\begin{center}
\unitlength 1.2mm
\begin{picture}(80,30)
\thicklines 
\put(10,20){\circle*{1}}
\put(30,20){\circle*{1}}
\put(50,20){\circle*{1}}
\put(60,20){\circle*{1}}
\put(70,20){\circle*{1}}
\put(20,10){\circle*{1}}
\put(40,10){\circle*{1}}
\put(60,10){\circle*{1}}
\put(10,20){\line(1,-1){10}}
\put(20,10){\line(1,1){10}}
\put(30,20){\line(1,-1){3}}
\put(40,10){\line(-1,1){3}}
\put(40,10){\line(1,1){10}}
\put(50,20){\line(1,-1){10}}
\put(60,10){\line(1,1){10}}
\put(60,10){\line(0,1){10}}
\put(35,15){\makebox(0,0){$\cdots$}}
\put(10,24){\makebox(0,0){$\id$}}
\put(30,24){\makebox(0,0){$\si$}}
\end{picture}
\end{center}
\caption{The principal graph for the subfactor
$D_{2n}$}
\label{D2n}
\end{figure}

Then we can easily verify Conditions 1, 2 (a) and 4.
We next verify Condition 3.  We label four irreducible
objects as in Figure \ref{D2n2}.  (If $n=3$, we set $\la_1=\id$.)
Note that the connection with respect to the generator
$\si$ has a principal graph as in Figure \ref{Deven}.
(See \cite{I1}, for example, for the fusion rules of
a subfactor with principal graph $D_{2n}$.)

\begin{figure}[htb]
\begin{center}
\unitlength 1.2mm
\begin{picture}(80,30)
\thicklines 
\put(10,20){\circle*{1}}
\put(30,20){\circle*{1}}
\put(50,20){\circle*{1}}
\put(60,20){\circle*{1}}
\put(70,20){\circle*{1}}
\put(20,10){\circle*{1}}
\put(40,10){\circle*{1}}
\put(60,10){\circle*{1}}
\put(10,20){\line(1,-1){10}}
\put(20,10){\line(1,1){3}}
\put(30,20){\line(-1,-1){3}}
\put(30,20){\line(1,-1){10}}
\put(40,10){\line(1,1){10}}
\put(50,20){\line(1,-1){10}}
\put(60,10){\line(1,1){10}}
\put(60,10){\line(0,1){10}}
\put(25,15){\makebox(0,0){$\cdots$}}
\put(10,24){\makebox(0,0){$\id$}}
\put(30,24){\makebox(0,0){$\la_1$}}
\put(50,24){\makebox(0,0){$\la_2$}}
\put(60,24){\makebox(0,0){$\la_3$}}
\put(70,24){\makebox(0,0){$\la_4$}}
\end{picture}
\end{center}
\caption{The principal graph for the subfactor
$D_{2n}$}
\label{D2n2}
\end{figure}

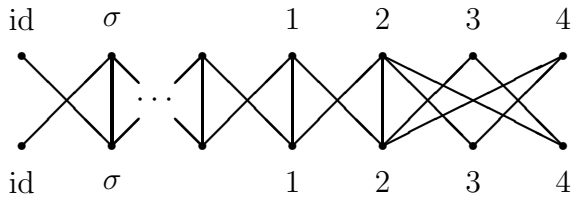
\begin{figure}[htb]
\begin{center}
\unitlength 1.2mm
\begin{picture}(80,30)
\thicklines 
\put(10,10){\circle*{1}}
\put(10,20){\circle*{1}}
\put(20,10){\circle*{1}}
\put(20,20){\circle*{1}}
\put(30,10){\circle*{1}}
\put(30,20){\circle*{1}}
\put(40,10){\circle*{1}}
\put(40,20){\circle*{1}}
\put(50,10){\circle*{1}}
\put(50,20){\circle*{1}}
\put(60,10){\circle*{1}}
\put(60,20){\circle*{1}}
\put(70,10){\circle*{1}}
\put(70,20){\circle*{1}}
\put(10,20){\line(1,-1){10}}
\put(10,10){\line(1,1){10}}
\put(20,10){\line(0,1){10}}
\put(20,10){\line(1,1){3}}
\put(30,20){\line(-1,-1){3}}
\put(20,20){\line(1,-1){3}}
\put(30,10){\line(-1,1){3}}
\put(30,20){\line(1,-1){10}}
\put(30,20){\line(0,-1){10}}
\put(30,10){\line(1,1){10}}
\put(40,10){\line(1,1){10}}
\put(40,10){\line(0,1){10}}
\put(40,20){\line(1,-1){10}}
\put(50,20){\line(1,-1){10}}
\put(50,20){\line(0,-1){10}}
\put(50,20){\line(2,-1){20}}
\put(50,10){\line(1,1){10}}
\put(50,10){\line(2,1){20}}
\put(60,10){\line(1,1){10}}
\put(60,20){\line(1,-1){10}}
\put(25,15){\makebox(0,0){$\cdots$}}
\put(10,24){\makebox(0,0){$\id$}}
\put(20,24){\makebox(0,0){$\sigma$}}
\put(40,24){\makebox(0,0){$1$}}
\put(50,24){\makebox(0,0){$2$}}
\put(60,24){\makebox(0,0){$3$}}
\put(70,24){\makebox(0,0){$4$}}
\put(10,6){\makebox(0,0){$\id$}}
\put(20,6){\makebox(0,0){$\sigma$}}
\put(40,6){\makebox(0,0){$1$}}
\put(50,6){\makebox(0,0){$2$}}
\put(60,6){\makebox(0,0){$3$}}
\put(70,6){\makebox(0,0){$4$}}
\end{picture}
\end{center}
\caption{The principal graph for the subfactor
$\sigma(M)\subset M$}
\label{Deven}
\end{figure}

We first claim that if the vertices $\la_3$ and $\la_4$ are not involved,
then the connection values with respect to the the generator
$\si$ are non-zero.  As in \cite[II, Section 3]{BE}, we may
assume that the irreducible objects of the tensor
category are realized as
$\{\a_0,\a_2,\dots,\a_{2n-4},\a_{2n-2}^{(1)},\a_{2n-2}^{(2)}\}$,
arising from $\a$-induction applied to the system
$SU(2)_{4n-4}$ having the irreducible objects
$\{0,1,2,\dots,4n-4\}$.  (Note that it does not matter whether we
use $\a^+$ or $\a^-$, so we have dropped the $\pm$ symbol.)
We denote, by $W(i,j,k,l)$, the connection value with respect
to the generator $\si=\a_2$ given by the square in
Figure \ref{conn}.
(Note that the value $W(i,j,k,l)$ depends on the choices of
intertwiners, but the absolute value $|W(i,j,k,l)|$ is
independent of such choices, since the intertwiner spaces are
now all one-dimensional.)
For example, assume $n>4$ and consider
the connection value $W(\a_4,\a_6,\a_4,\a_6)$.
By \cite[II, Section 3]{BE}, all the four intertwiners involved
in this connection come from the intertwiners for
$SU(2)_{4n-4}$, and thus the connection value is given
by the connection $W(4,6,4,6)$ for $SU(2)_{4n-4}^{\rm even}$
with respect to the generator $2$.  This value is given
as a single term of $6j$-symbols of $SU(2)_{4n-4}$ and it
is non-zero by \cite{KLi}.  The general case is dealt with
in the same method.

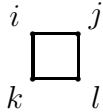
\begin{figure}[htb]
\begin{center}
\unitlength 0.6mm
\begin{picture}(30,30)
\thicklines 
\put(10,10){\circle*{1}}
\put(10,20){\circle*{1}}
\put(20,10){\circle*{1}}
\put(20,20){\circle*{1}}
\put(10,10){\line(1,0){10}}
\put(10,10){\line(0,1){10}}
\put(20,20){\line(-1,0){10}}
\put(20,20){\line(0,-1){10}}
\put(6,24){\makebox(0,0){$i$}}
\put(24,24){\makebox(0,0){$j$}}
\put(6,6){\makebox(0,0){$k$}}
\put(24,6){\makebox(0,0){$l$}}
\end{picture}
\end{center}
\caption{A connection value for $D_{2n}^{\rm even}$}
\label{conn}
\end{figure}

Thus, we consider the remaining case where all the four vertices
of the connection value are one of $\la_1,\la_2,\la_3,\la_4$.
In the below, we denote the vertices
$\la_1,\la_2,\la_3,\la_4$ simply by $1,2,3,4$.
Denote this statistical dimensions of $1,2,3,4$ by
$d_1,d_2,d_3,d_4$ respectively.  Their explicit
values are as follows.
\begin{eqnarray}
\label{ds}
d_1 &=& \frac{\sin \frac{2n-5}{4n-2}\pi}{\sin\frac{\pi}{4n-2}},\nonumber\\
d_2 &=& \frac{\sin \frac{2n-3}{4n-2}\pi}{\sin\frac{\pi}{4n-2}},\\
d_3 &=& d_4 = \frac{1}{2\sin\frac{\pi}{4n-2}}.\nonumber
\end{eqnarray}
For a fixed pair $(i,l)$, we denote the unitary matrix 
$(W(i,j,k,l))_{j,k}$ by $W_{il}$.  Using the bi-unitarity Axioms 1 and
4 in \cite[Chapter 10]{EK}, originally due to \cite{O1},
we compute several matrices $W_{il}$ below.
Recall that the renormalization Axiom 4 in \cite[Chapter 10]{EK}
now implies
$$|W(i,j,k,l)|=\sqrt{\frac{d_j d_k}{d_i d_l}}|W(j,i,l,k)|.$$

If $i=1$ and $l\neq 3,4$, then the entries in $W_{il}$
are again given as single terms of the $6j$-symbols of
$SU(2)_{4n-4}$ and thus, they are non-zero.
The unitary matrices $W_{13}$ and $W_{14}$ have size
$1\times 1$, so the entries are obviously non-zero.

The unitary matrix $W_{21}$ has a size $2\times 2$,
and all the entries in $W_{il}$
are again given as single terms of the $6j$-symbols of
$SU(2)_{4n-4}$ and thus, they are non-zero.

The unitary matrix $W_{22}$ has a size $4\times 4$.
The entry $W(2,1,1,2)$ is non-zero because we have already
seen that $W_{11}$ has no zero entries and we have the
renormalization axiom.  Similarly, the entries
$W(2,2,1,2)$, $W(2,1,2,2)$,
$W(2,3,1,2)$, $W(2,1,3,2)$,
$W(2,4,1,2)$, and $W(2,1,4,2)$ are non-zero.

The entry $W(2,2,2,2)$ is also given as a
single term of the $6j$-symbols of
$SU(2)_{4n-4}$ and thus, it is non-zero.

We assume $W(2,3,2,2)=0$ and will derive a contradiction.
Using the renormalization axiom twice, we obtain
$W(2,2,3,2)=0$.  Another use of the renormalization axiom
gives $W(3,2,2,2)=0$.  Since the $2\times2$ matrix
$W_{32}$ is unitary, this implies $|W(3,4,2,2)|=1$.
The renormalization axiom then gives $|W(2,2,3,4)|=1$.
Since the $2\times2$ matrix $W_{24}$ is unitary, this
gives $W(2,3,3,4)=W(2,2,2,4)=0$. 
These two equalities then give
$W(3,4,2,3)=0$ and $W(2,2,4,2)=0$
with the renormalization axiom, respectively.
Thus we have verified the $(2,4)$-entry of the $4\times4$
unitary matrix $W_{22}$ is zero.  Similarly, its $(4,2)$-entry
is also zero.  The identity $W(3,4,2,3)=0$ and unitarity of
the $2\times2$ matrix $W_{33}$ give $|W(3,2,2,3)|=1$.  The
renormalization axiom then produces
$|W(2,3,3,2)|=d_3/d_2$.  The $1\times 1$ matrix $W_{43}$ is
unitary, thus the renormalization axiom gives
$|W(2,4,3,2)|=d_3/d_2$.  Similarly, we obtain
$|W(2,3,4,2)|=d_3/d_2$.  The $1\times 1$ matrix $W_{13}$ is
unitary, thus the renormalization axiom gives
$|W(2,1,3,2)|=\sqrt{d_1 d_3}/d_2$.  Now we use the orthogonality
of the second and third row vectors of the $4\times 4$ unitary
matrix $W_{22}$.  We have so far obtained that the
$(2,3)$, $(2,4)$, $(3,2)$-entries are zero and the $(3,1)$-entry
is non-zero.  We thus know that the $(2,1)$-entry is zero, but
this is a contradiction because we have already seen above that
the $(2,1)$-entry $W(2,1,2,2)$ is non-zero.
We have thus proved $W(2,3,2,2)\neq0$.
By a similar method, we can prove that
$W(2,4,2,2)$, $W(2,2,3,2)$ and $W(2,2,4,2)$ are all non-zero.

We next assume $W(2,3,3,2)=0$.
For the same reason as above, we obtain
\begin{eqnarray}
\label{4by41}
|W(2,3,1,2)|&=&|W(2,4,1,2)|=|W(2,1,3,2)|=|W(2,1,4,2)|=
\frac{\sqrt{d_1 d_3}}{d_2},\\
\label{4by42}
|W(2,3,4,2)|&=&|W(2,4,3,2)|=\frac{d_3}{d_2}.
\end{eqnarray}
Since $W(2,3,3,2)=0$, the renormalization axiom implies
$W(3,2,2,3)=0$.  Since the $2\times 2$-matrix
$W_{33}$ is unitary, we obtain $|W(3,2,4,3)|=1$.
The renormalization axiom gives 
$|W(2,3,3,4)|=\sqrt{d_3/d_2}$.  Unitarity of the
$2\times 2$-matrix $W_{24}$ then gives
$|W(2,2,2,4)|=\sqrt{d_3/d_2}$, which then gives
$|W(2,2,4,2)|=|W(2,4,2,2)|=d_3/d_2$ with the renormalization 
axiom.  The identities (\ref{ds}), together with a simple
computation of trigonometric functions, give
\begin{equation}\label{norm1}
d_1 d_3+ 2d_3^2 = d_2^2.
\end{equation}
Since the third row vector, the fourth row vector, and the third
column vector of the unitary matrix $W_{22}$ have a norm 1,
this identity (\ref{norm1}), together with (\ref{4by41}), 
(\ref{4by42}) gives
$|W(2,2,3,2)|=|W(2,3,2,2)|=d_3/d_2$ and $W(2,4,4,2)=0$.
Thus the matrix $A=(A_{jk})_{jk}=(|W(2,k,j,2)|)_{jk}$ is given
as follows, where $\a,\be,\ga$ are non-negative real numbers.
\begin{equation}\label{matrix}
\left(
\begin{array}{cccc}
\a & \be & \displaystyle\frac{\sqrt{d_1 d_3}}{d_2} &
\displaystyle\frac{\sqrt{d_1 d_3}}{d_2}\\
\be & \ga & \displaystyle\frac{d_3}{d_2} &
\displaystyle\frac{d_3}{d_2} \\
\displaystyle\frac{\sqrt{d_1 d_3}}{d_2} &
\displaystyle\frac{d_3}{d_2} & 0 & \displaystyle\frac{d_3}{d_2} \\
\displaystyle\frac{\sqrt{d_1 d_3}}{d_2} &
\displaystyle\frac{d_3}{d_2} & \displaystyle\frac{d_3}{d_2} & 0
\end{array}
\right)
\end{equation}
Orthogonality of the first and third row vectors of
$W_{22}$ implies
\begin{equation}\label{orth}
\frac{\sqrt{d_1 d_3} d_3}{d_2^2} \leqq
\a \frac{\sqrt{d_1 d_3}}{d_2} + 
\be \frac{d_3}{d_2}.
\end{equation}
Since the first row vector of $W_{22}$ has a norm 1, we also
have
\begin{equation}\label{norm}
\a^2+\be^2=1-\frac{2d_1 d_3}{d_2^2}.
\end{equation}
The Cauchy-Schwarz inequality with (\ref{orth}), (\ref{norm}),
we obtain
$$\frac{\sqrt{d_1} d_3}{d_2}\leqq
\sqrt{d_1+d_3}\sqrt{1-\frac{2d_1 d_3}{d_2^2}},$$
which, together with (\ref{norm1}), implies
$$\sqrt{d_1} d_3 \leqq \sqrt{d_1+d_3}\sqrt{2d_3^2-d_1 d_3}.$$
This implies $d_1^2\leqq 2d_3$, which gives
\begin{equation}\label{trig}
\sin^2 \frac{2n-5}{4n-2} \pi \leqq \frac{1}{2}
\end{equation}
by (\ref{ds}).  This inequality (\ref{trig}) fails, if
we have $(2n-5)/(4n-2) > 1/4$, that is, $n>9/2$.  Since we
now assume $n\ge 3$, this has produced a contradiction
and we have shown $W(2,3,3,2)\neq0$,
unless $n=3, 4$.
We deal with the remaining two cases $n=3, 4$ by direct
computations of the connection as follows.

If $n=3$, we have the Dynkin diagram $D_6$.  A subfactor
with principal with $D_6$ is realized as the asymptotic
inclusion \cite[page 137]{O1}, \cite[Definition 12.23]{EK},
\cite[Section 2]{I3}, of a subfactor with principal graph $A_4$
as in \cite[Section III.1]{O2}, \cite[page 663]{EK},
\cite[Theorem 4.1]{I3}.  Thus the tensor category
$D_6^{\rm even}$ is realized as a self-tensor product of
the tensor category of $A_4^{\rm even}$ and that our
current generator $\si$ is realized as a tensor product of
the standard generators in two copies of $A_4^{\rm even}$.
As in Case 2 (b) below, the connection values
are non-zero for $A_4^{\rm even}$, thus our current connection
values are also non-zero as products of two non-zero values.

We finally deal with the case $n=4$.  We label the even
vertices of the principal graph $D_8$ as in Figure \ref{D8}.

\begin{figure}[htb]
\begin{center}
\unitlength 1.2mm
\begin{picture}(80,30)
\thicklines 
\put(10,20){\circle*{1}}
\put(30,20){\circle*{1}}
\put(50,20){\circle*{1}}
\put(60,20){\circle*{1}}
\put(70,20){\circle*{1}}
\put(20,10){\circle*{1}}
\put(40,10){\circle*{1}}
\put(60,10){\circle*{1}}
\put(10,20){\line(1,-1){10}}
\put(20,10){\line(1,1){10}}
\put(30,20){\line(1,-1){10}}
\put(40,10){\line(1,1){10}}
\put(50,20){\line(1,-1){10}}
\put(60,10){\line(1,1){10}}
\put(60,10){\line(0,1){10}}
\put(10,24){\makebox(0,0){$0$}}
\put(30,24){\makebox(0,0){$1$}}
\put(50,24){\makebox(0,0){$2$}}
\put(60,24){\makebox(0,0){$3$}}
\put(70,24){\makebox(0,0){$4$}}
\end{picture}
\end{center}
\caption{The principal graph for the subfactor
$D_8$}
\label{D8}
\end{figure}
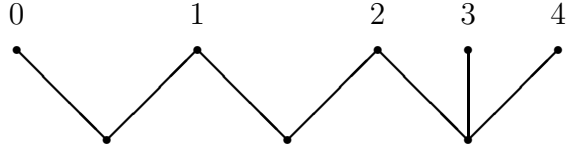

We continue the
computations of $|W(i,j,k,l)|$'s using the matrix (\ref{matrix}),
where the non-negative real numbers $\a,\be,\ga$ have been
defined.  The renormalization axiom gives
$|W(1,1,2,1)|=\sqrt{d_2/d_1}|W(1,1,1,2)|$ and unitarity of
the $2\times2$-matrix $W_{12}$ gives $|W(1,2,2,2)|=|W(1,1,1,2)|$.
So we have
\begin{equation}\label{tildebeta}
|W(1,1,2,1)|=|W(1,2,1,1)|=\sqrt{\frac{d_2}{d_1}}|
W(1,2,2,2)|=|W(2,1,2,2)|=\be,
\end{equation}
again by the renormalization.  We also have
\begin{equation}\label{tildebeta2}
|W(1,2,1,1)|=\be.
\end{equation}
Unitarity of the $1\times1$-matrix $W_{02}$ gives
$|W(0,1,1,2)|=1$ and thus, the renormalization axiom gives
\begin{equation}\label{corner}
|W(1,0,2,1)|=|W(1,2,0,1)|=\frac{\sqrt{d_2}}{d_1},
\end{equation}
since $d_0=1$.  Similarly, unitarity of the $1\times1$-matrix
$W_{01}$ gives
\begin{equation}\label{corner2}
|W(1,0,1,1)|=|W(1,1,0,1)|=\frac{1}{\sqrt{d_1}},
\end{equation}
and unitarity of the $1\times1$-matrix
$W_{00}$ gives
\begin{equation}\label{corner3}
|W(1,0,0,1)|=\frac{1}{d_1}.
\end{equation}
We also have
\begin{equation}\label{tildealpha}
|W(1,2,2,1)|=\frac{d_2}{d_1} |W(2,1,1,2)|=\frac{d_2}{d_1}\a,
\end{equation}
Thus the $3\times3$-matrix $B=(B_{jk})_{jk}=(|W(1,k,j,1)|)_{jk}$
is given as follows, where $\de$ is a non-negative real number,
by (\ref{tildebeta}), (\ref{tildebeta2}), (\ref{corner}),
(\ref{corner2}), (\ref{corner3}), (\ref{tildealpha}).
\begin{equation}\label{matrix2}
\left(
\begin{array}{ccc}
\displaystyle\frac{1}{d_1} & \displaystyle\frac{1}{\sqrt{d_1}} &
\displaystyle\frac{\sqrt{d_2}}{d_1}  \\
\displaystyle\frac{1}{\sqrt{d_1}} & \de & \be \\
\displaystyle\frac{\sqrt{d_2}}{d_1} & \be &
\displaystyle\frac{d_2}{d_1}\a
\end{array}
\right)
\end{equation}
The first row vector of the matrix (\ref{matrix}) has a norm 1,
thus we have
\begin{equation}\label{norm2}
\a^2+\be^2=1-\frac{2d_1d_3}{d_2^2}.
\end{equation}
The third row vector of the matrix (\ref{matrix2}) has a norm 1,
thus we have
\begin{equation}\label{norm3}
\frac{d_2}{d_1^2}+\be^2+\frac{d_2^2}{d_1^2}\a^2=1.
\end{equation}
Equations (\ref{norm2}) and (\ref{norm3}) give the following
value for $\be^2$.
\begin{equation}\label{beta2}
\be^2=\frac{d_2^2-2d_1d_3-d_1^2+d_2}{d_2^2-d_1^2}.
\end{equation}
Note that the denominator is not zero.
Let $t$ be the index of the subfactor with principal graph
$D_8$.  (That is, $t=4\cos^2 \pi/14$.)  Then the Perron-Frobenius
theory gives the following identities.
\begin{eqnarray*}
d_1&=& t-1,\\
d_2&=& t^2-3t+1,\\
d_3&=& \frac{t^3-5t^2+6t-1}{2}.
\end{eqnarray*}
Then these imply $d_2^2-2d_1d_3-d_1^2+d_2=0$ in (\ref{beta2}),
we thus obtain $\be=0$, which has been already excluded above.
We have thus reached a contradiction and shown $W(2,3,3,2)\neq0$.

Similarly, we can prove $W(2,4,4,2)\neq0$.

The unitary matrix $W_{34}$ has a size $1\times1$, so the
renormalization axiom implies $W(2,4,3,2)\neq0$.  Similarly,
we have $W(2,3,4,2)\neq0$.  We have thus proved that all the
entries of $W_{22}$ are non-zero.

The unitary matrix $W_{23}$ has a size $2\times 2$.
If this matrix has a zero entry, we have
either $W(2,2,2,3)=W(2,4,4,3)=0$ or
$W(2,2,4,3)=W(2,4,2,3)=0$.  The former case,
together with the renormalization axiom, implies
$W(2,2,3,2)=0$, which is already excluded in the above
study of $W_{22}$.  The latter case gives 
$|W(2,4,4,3)|=1$, which, 
together with the renormalization axiom, implies
$|W(4,2,3,4)|=\sqrt{d_2/d_4}>1$ by (\ref{ds}).  This is against the
unitarity axiom and thus cannot happen.

The $2\times 2$ unitary matrix $W_{24}$ is dealt with
in a similar way to the case $W_{23}$.

The unitary matrices $W_{31}$ and $W_{34}$ also
have size $1\times 1$, so the entries are again non-zero.
The matrices $W_{32}$ and $W_{33}$ have size $2\times 2$.
The entries of $W_{32}$ have the same absolute values as
the entries of $W_{23}$, so the above arguments for $W_{23}$
show that they are non-zero.
We next consider $W_{33}$.  If this $2\times 2$ unitary matrix
contains a zero entry, then we have
either $W(3,2,2,3)=W(3,4,4,3)=0$ or $W(3,2,4,3)=W(3,4,2,3)=0$.  
The former case, together with the renormalization axiom, implies
$W(2,3,3,2)=0$, which is already excluded in the above
study of $W_{22}$.  
The latter case, together with the renormalization axiom, implies
$W(2,3,3,4)=0$, which is already excluded in the above
study of $W_{24}$.

The four matrices $W_{4l}$ can be dealt with in the same
way as above for $W_{3l}$.  

Thus we are done for Case 1 (b).

Case 1 (c).  Only fusion rules and $6j$-symbols matter, and the
braiding does not matter, for the Conditions
in Theorem \ref{vanishing}, so our tensor
category can be identified with
$SU(2)_2$ and this is a special case of Case 1 above.

Case 1 (d).  In a similar way to the above case,
this tensor category can be identified with
the even part of the tensor category $SU(2)_3$, 
so this is a special case of Case 2 (b) below.

Case 2 (a).  This is trivial.

Case 2 (b).  We label the irreducible objects of the 
tensor category
$SU(2)_k$ with index $0,1,2,\dots, k$, as above.  (We also use
the name $A_{k+1}^{\rm even}$ for this tensor category.)
Let $\si$ be the
generator $2$ this time.  Conditions 1 and 2 (a) of
Theorem \ref{vanishing} clearly hold.
Since all $6j$-symbols for $SU(2)_k$ have non-zero
values as in \cite{KLi}, Condition 3 holds, in particular.
The multiplication rule by the generator $\si$
is described with the even steps of the
usual Bratteli diagram for the principal
graph $A_{k+1}$ as in \cite{J}, \cite[Chapter 9]{EK}, so we see
that Condition 4 holds.

Case 3 (a). This is the Virasoro tensor category with
central charge $c=1-6/n(n+1)$.  We recall the description of
the irreducible objects in the tensor
category given by \cite[Theorem 4.6]{X2}
applied to $SU(2)_{n-1}\subset SU(2)_{n-2}\otimes SU(2)_1$, as follows.
(Also see \cite[Section 3]{KL} for our notations.)
We now have a net of subfactors $\Vir_c \otimes SU(2)_{n-1}
\subset SU(2)_{n-2}\otimes SU(2)_1$ with finite index and apply
the $\a$-induction to this inclusion.
The irreducible representations of the net $\Vir_c$ are labeled as
$$\{\si_{j,k} \mid j=0,1,\dots,n-2, \quad k=0,1,\dots,n-1,\quad
j+k\in 2\Z\}.$$
Xu's result \cite[Theorem 4.6]{X2} then shows the following.
First, the systems
$\{\si_{j,k}\}$ and  $\{\a_{\si_{j,k}\otimes\id}\}$ have the
isomorphic fusion rules and $6j$-symbols.  Furthermore, the latter
system is isomorphic to the system
$$\{(\la'_j\otimes\id)(\a_{\id\times\la_k})\mid 
j=0,1,\dots,n-2, \quad k=0,1,\dots,n-1,\quad
j+k\in 2\Z\},$$
where $\{\la_k\mid k=0,1,\dots,n-1\}$ and
$\{\la'_j\mid j=0,1,\dots,n-2\}$ are the system of irreducible DHR
endomorphisms of the nets $SU(2)_{n-1}$ and $SU(2)_{n-2}$, respectively.
This system have further isomorphic fusion rules and $6j$-symbols
to the system
\begin{equation}\label{product}
\{\la'_j\otimes \la_k\mid j=0,1,\dots,n-2,
\quad k=0,1,\dots,n-1,\quad j+k\in 2\Z\},
\end{equation}
of irreducible DHR endomorphisms of the net
$SU(2)_{n-2}\otimes SU(2)_{n-1}$.  (Note that we have a restriction
$j+k\in2\Z$, so this system is a subsystem of that of the all the
irreducible DHR endomorphisms of the net
$SU(2)_{n-2}\otimes SU(2)_{n-1}$.)
As in \cite[Section 3]{KL}, we can identify the system of
these $\si_{j,k}$'s with the system of
characters of the minimal models \cite[Subsection 7.3.4]{DMS}
whose fusion rules are given in \cite[Subsection 7.3.3]{DMS}.
We take the DHR endomorphism $\si_{1,1}$ as $\si$ in 
Theorem \ref{vanishing} and then,
from these fusion rules, we easily see that Condition 1 holds.
It is also easy to see that we have a natural $\Z/2\Z$-grading such
that $\si$ is an odd generator, so Condition 2 (b) holds.
By considering the connection of the system (\ref{product}),
we know that the connection value with respect to the generator $\si$
is a product of the two connection values
of the systems $SU(2)_{n-2}$
and $SU(2)_{n-1}$  with respect to the standard generators.
Since these two connection values for 
$SU(2)_{n-2}$ and $SU(2)_{n-1}$ are the usual
connection values for the paragroups labeled with the Dynkin
diagrams $A_{n-1}$ and $A_n$, and they are non-zero by \cite{O1},
\cite{K1}, \cite[Section 11.5]{EK}, we conclude that Condition 3 
holds.  From the fusion rule described as above, we verify that 
Condition 4 also holds.  (Recall the comment on Condition 4 after
the statement of Theorem \ref{vanishing} and draw the principal 
graph for a subfactor given by $\si_{1,1}$.)

Case 3 (b).  The tensor category is produced with $\a$-induction and a
simple current extension of index 2 as in \cite[II, Section 3]{BE}.
The fusion rules and $6j$-symbols are given by a direct product of
the two systems $A_{4n}^{\rm even}$ and $D_{2n+2}^{\rm even}$.
We can use the direct product of the $\si$ in Figure \ref{D2n} and
the $\si$ in Case 2 (b) as the current $\si$ for
Theorem \ref{vanishing}.  Then Conditions 1, 2 (a),
and 4 easily follow and the connection values are non-zero as
products of non-zero values in Cases 1 (b) and 2.

Case 3 (c).  This case is proved in a similar way to the above
proof of case 3 (b).

Case 3 (d).  The tensor category is produced with $\a$-induction as
in \cite[Section 4.2]{KL}.
The irreducible objects of the tensor category are labeled
with pairs $(j,k)$ with $j=0,1,\dots,9$ and $k=0,1,2$ with
$j+k\in 2\Z$.  The fusion rules
of the objects $\{(j,0)\mid j=0,1,\dots,9\}$ obey the $A_{10}$
fusion rule and those of $\{(0,0), (0,1), (0,2)\}$ obey the
$A_3$ fusion rule.  Let $\si$ be the object $(1,1)$.  Then as in
Case 1, we can verify Conditions 1, 2 (b), 3 and 4.

Case 3 (e).  This case is proved in a similar way to the above
proof of case 3 (d).

Case 3 (f).  The tensor category is again produced with 
$\a$-induction as in \cite[Section 4.2]{KL}.
The fusion rules and $6j$-symbols are given as the direct product of
the two systems $A_{28}^{\rm even}$ and $A_4^{\rm even}$.  The irreducible
objects of the former system are labeled with $0,2,\dots,26$
as usual, and the latter system is given as $\{\id, \tau\}$ with
$\tau^2=\id\oplus\tau$.  Then we can choose $(14,\tau)$ as $\si$ and
verify Conditions 1, 2 (a), 3 and 4, using the same arguments as in
Cases 2 and 3 (b).

Case 3 (g).  This case is proved in a similar way to the above
proof of case 3 (f).

Case 4 (a).  Now, the only non-trivial sub-tensor categories are
$\Z/2\Z$, $SU(2)_{n-2}$, $SU(2)_{n-2}^{\rm even}$,
$SU(2)_{n-1}$, $SU(2)_{n-1}^{\rm even}$ and the even parts 
with respect to the
$\Z/2\Z$-grading described in the above proof of Case 3 (a).
The conclusion trivially holds for the first case.
The next four cases have been already dealt with in
Cases 1(a) and 2 (b).
In the last case, we can identify the tensor category with the
direct product of two tensor categories $SU(2)_{n-2}^{\rm even}$
and $SU(2)_{n-1}^{\rm even}$.  We use the same labeling of
the irreducible DHR sectors as in the proof of the Case
3 (a) and then we can use the generator $\si_{2,2}$ as
$\si$ in Theorem \ref{vanishing}. 

Case 4 (b).  The only non-trivial sub-tensor
categories we have are now
$A_{4n}^{\rm even}$ and $D_{2n+2}^{\rm even}$.  Thus, we have
the conclusion by Cases 2 (b) and 1 (b), respectively.

Case 4 (c).  This case is proved in a similar way to the above
proof of case 4 (b).

Case 4 (d).  The only non-trivial sub-tensor 
categories we have are now
$\Z/2\Z$, $A_{10}^{\rm even}$, their direct product,
and $A_3$.
We can deal with the group $\Z/2\Z$ trivially.  The cases
$A_{10}^{\rm even}$ and $A_3$ are particular cases of Cases 2 (b)
and 1 (a), respectively.
For the case of the direct product of $A_{10}^{\rm even}$ and
$\Z/2\Z$, we can choose $\si=(2,2)$ in the notation of the proof
for Case 3 (d).

Case 4 (e).  This case is proved in a similar way to the above
proof of case 4 (d).

Case 4 (f).  The only non-trivial sub-tensor categories we have are now
$A_4^{\rm even}$ and $A_{28}^{\rm even}$.  Both are
special cases of Case 2 (b).

Case 4 (g).  This case is proved in a similar way to the above
proof of case 4 (f).
\end{proof}

\begin{remark}{\rm We have the following application of the
above theorem.
Consider the tensor category corresponding to the WZW-model
$SU(2)_{28}$.  Regard the irreducible objects as irreducible
endomorphisms of a type III factor $M$ and label them as
$\id=\la_0,\la_1,\la_2,\dots,\la_{28}$ as usual.
Then the endomorphism
$\ga=\la_0\oplus\la_{10}\oplus\la_{18}\oplus\la_{28}$ is
a dual canonical endomorphism and uniqueness of
$Q$-system $(\ga,V,W)$ up to unitary equivalence was shown
in \cite[Section 6]{KO} based on a result in vertex operator
algebras.  (This uniqueness was used in our previous work \cite{KL}.)
Izumi has also given another proof of this uniqueness with a more
direct method.  We remark that our above theorem also gives a
different proof of this uniqueness as follows.

We may assume that $M$ is injective.
Suppose that we have two endomorphisms $\rho_1,\rho_2$ of $M$
such that $\rho_1\bar\rho_1=\rho_2\bar\rho_2=\ga$.
As in \cite[Proposition A.3]{BEK2}, we can prove that
the two subfactors $\rho_1(M)\subset M$ and $\rho_2(M)\subset M$
have the isomorphic higher relative commutants, and then we 
conclude by \cite[Corollary 6.4]{P2} that the two subfactors
are isomorphic via $\th\in\Aut(M)$.  We then may and do assume
$\rho_2=\th\cdot\rho_1$ and now we have 
$\th\cdot\ga\cdot\th^{-1}=\ga$.
Since $\ga=\la_0\oplus\la_{10}\oplus\la_{18}\oplus\la_{28}$
and powers of $\ga$ produce all of 
$\la_0,\la_2,\la_4,\dots,\la_{28}$, we know that
$[\th\cdot\la_{2j}\cdot\th^{-1}]=[\la_{2j}]$ for
$j=0,1,2,\dots,14$, where the sqare brackets denote the
unitary equivalence classes.
Then we have a map
$$\th: \Hom(\la,\mu)\ni t\mapsto \th(t)\in
\Hom(\th\cdot\la\cdot\th^{-1},\th\cdot\mu\cdot\th^{-1})$$
giving an automorphism of the tensor category generated by
powers of $\ga$.
By Case 2 of Theorem \ref{SU2-Virasoro}, this automorphism
$\th$ is trivial in the sense of
Definition \ref{vanish}.  The automorphism $\th$ sends the
$Q$-system $(\ga, V_1, W_1)$ for $\rho_1$ to the one
$(\ga, V_2, W_2)$ for $\rho_2$, and now the triviality of
$\th$ implies that these two systems are unitarily equivalent.
}\end{remark}

Using the above Theorem \ref{SU2-Virasoro}, 
we obtain the following classification result of
2-dimensional completely rational nets.  The 
meaning of the condition that the $\mu$-index is 1 will be
further studied in the next section.

Consider a 2-dimensional local
completely rational conformal net $\B$ with 
central charge $c=1-6/m(m+1)<1$ and $\mu$-index $\mu_\B=1$.
By \cite{R2}, we have inclusions
$$\A_L \otimes \A_R \subset 
\A_L^{\max} \otimes \A_R^{\max} \subset \B,$$
where $\A_L, A_R, \A_L^{\max}, \A_R^{\max}$ are one-dimensional 
local conformal
nets.  By assumption, $\A_L^{\max}$ and $\A_R^{\max}$ have the same central
charge $c$.  Rehren's result \cite[Corollary 3.5]{R2} and
our results \cite[Proposition 24]{KLM} together imply that
the fusion rules of the systems of entire irreducible DHR endomorphisms
of the two nets $\A_L^{\max}, \A_R^{\max}$ are
isomorphic, and our previous result \cite[Theorem 5.1]{KL} implies
that the two nets  $\A_L^{\max}, \A_R^{\max}$ are isomorphic as nets.
Since both $\A_L^{\max}, \A_R^{\max}$ contain $\Vir_c$ as subnets,
we obtain an irreducible inclusion $\Vir_c\otimes \Vir_c \subset \B$.
A decomposition of a vacuum sector of $\B$ restricted on 
$\Vir_c\otimes \Vir_c$ produces a decomposition matrix 
$(Z_{\la\mu})_{\la\mu}$, where $\la,\mu$ are representatives
of uintary equivalence classes of irreducible DHR
endomorphisms of the net $\Vir_c$.  Since $\mu_\B=1$,
by Theorem \ref{modular}, due to M\"uger \cite{Mu}, we know that this
matrix $Z$ is a modular invariant of 
the Virasoro tensor category $\Vir_c$ and such 
modular invariants have been classified by
Cappelli-Itzykson-Zuber \cite{CIZ} as in Table \ref{CIZ-mod}.

\begin{table}[htbp]
\begin{center}
\begin{tabular}{|c|c|c|}\hline
$m$ & Labels for modular invariants in \cite{CIZ} & Type \\ \hline
$n$ & $(A_{n-1}, A_n)$ & I\\ \hline
$4n$ & $(D_{2n+1}, A_{4n})$ & II\\ \hline
$4n+1$ & $(A_{4n}, D_{2n+2})$ & I \\ \hline
$4n+2$ & $(D_{2n+2}, A_{4n+2})$ & I\\ \hline
$4n+3$ & $(A_{4n+2}, D_{2n+3})$ & II \\ \hline
$11$ & $(A_{10}, E_6)$ & I \\ \hline
$12$ & $(E_6, A_{12})$ & I \\ \hline
$17$ & $(A_{16}, E_7)$ & II \\ \hline
$18$ & $(E_7, A_{18})$ & II \\ \hline
$29$ & $(A_{28}, E_8)$ & I \\ \hline
$30$ & $(E_8, A_{30})$ & I \\ \hline
\end{tabular}
\caption{Modular invariants for the Virasoro tensor category $\Vir_c$}
\label{CIZ-mod}
\end{center}
\end{table}

We claim that this correspondence from $\B$ to $Z$ 
is bijective.

\begin{theorem}\label{classif}
The above correspondence from $\B$ to $Z$ gives a bijection from
the set of isomorphism classes of such two-dimensional nets to
the set of modular invariants $Z$ in Table \ref{CIZ-mod}.
\end{theorem}

\begin{proof}
We first prove that this correspondence is surjective.
Take a modular invariant $Z$ in Table \ref{CIZ-mod}.
By \cite[Subsections 4.1, 4.2, 4.3]{KL}, we conclude that
this modular invariant can be realized
with $\a$-induction as in \cite[Corollary 5.8]{BEK1} for
extensions of the Virasoro nets.  Then
Rehren's results in \cite[Theorem 1.4, Proposition 1.5]{R3} imply that
we have a corresponding $Q$-system and a local extension
$\B\supset \Vir_c\otimes \Vir_c$ and that this $\B$ produces the
matrix $Z$ in the above correspondence.

We next show injectivity of the map.  Suppose that we have inclusion
$$\A_L \otimes \A_R \subset 
\A_L^{\max} \otimes \A_R^{\max} \subset \B,$$
where $\A_L, \A_R$ are isomorphic to $\Vir_c$ and that
this decomposition gives a matrix $Z$.  We have to prove that
the net $\B$ is uniquely determined up to isomorphism.
Recall that the nets $\A_L^{\max}$ and $\A_R^{\max}$ are among
those classified by \cite[Theorem 5.1]{KL}.
As we have seen above, $\A_L^{\max}$ and $\A_R^{\max}$ are
isomorphic as nets and we can naturally identify them.  
This isomorphism class and an isomorphism
$\pi$ from a fusion rule of $\A_L^{\max}$ onto that
of $\A_R^{\max}$ are uniquely determined by $Z$ by
\cite[Theorem 5.1]{KL}  (Also see \cite{BE4}.)

If the modular invariant is of type I, then we can naturally
identify $\A_L^{\max}$ and $\A_R^{\max}$ and the map $\pi$
is trivial.  Then the $Q$-system for 
the inclusion $\A_L^{\max} \otimes \A_R^{\max} \subset \B$
has a standard dual canonical endomorphisms as in the
Longo-Rehren $Q$-system and
the above results Corollary \ref{coro}, Theorems \ref{vanishing},
\ref{SU2-Virasoro} imply that this $Q$-system is equivalent
to the Longo-Rehren $Q$-system.

If the modular invariant is of type II, then we have
a non-trivial fusion rule automorphism $\pi$.  We then know
by \cite[Lemma 5.3]{BE4} that this fusion rule automorphism $\pi$
actually gives an automorphism of the tensor category
acting non-trivially on irreducible objects.
The same arguments as in the proof of
Theorem \ref{Q-sys} show that 2-cohomology vanishing implies
uniqueness of the $Q$-system.  Again, 
the above results Corollary \ref{coro}, Theorems \ref{vanishing},
\ref{SU2-Virasoro} give the 2-cohomology vanishing, thus
we have the desired uniqueness of the $Q$-system for the
inclusion $\A_L^{\max} \otimes \A_R^{\max} \subset \B$.
\end{proof}

\begin{remark}{\rm
In the case the modular invariant $Z$ above is of type II,
the automorphism $\pi$ of the tensor category above is actually
an autmorphism of a braided tensor category, as seen from the
above proof.

In the above classification, we have shown 2-cohomology vanishing
without assuming locality.  In the context of classification
of two-dimensional nets, this means that any (relatively
local irreducible) extension $\B$ of $\A_L^{\max} \otimes \A_R^{\max}$
with $\mu$-index being 1 is automatically local.
}\end{remark}
\section{The $\mu$-index, maximality of extensions, and
classification of non-maximal nets}

In Theorem \ref{classif}, we have classified 2-dimensional
completely rational local conformal nets and
central charge less than 1 under the assumption that the $\mu$-index
is 1.  In this section, we clarify the meaning of this condition
on the $\mu$-index.  As we have seen above, this condition is
equivalent to triviality of the superselection structure of the
net.  We further show that this condition is equivalent to
maximality of extensions of the 2-dimensional net, when we have a parity
symmetry for the net $\B$.  Here the net $\B$ is said to have a
parity symmetry if we have a
vacuum-fixing unitary involution $P$ such that  $P \B(O) P = \B(pO)$,
where $p$ maps $x+t \mapsto x-t$ in the two-dimensional Minkowski space.
In this case, $P$ clearly implements an isomorphism of
$\A_L$ and $\A_R$ and thus, an isomorphism of
$\A_L^{\max}$ and $\A_R^{\max}$.

Suppose we have a local extension $\cC$ of the two-dimensional
completely rational local conformal net $\B$ and the inclusion
$\B\subset \cC$ is strict.  Then we have $\mu_\B > \mu_C \ge 1$
by \cite[Proposition 24]{KLM}.  That is, if the net $\B$ is
not maximal with respect to local extensions, then we have
$\mu_\B > 1$.  This argument
does not require a parity symmetry condition.

Conversely, suppose we have $\mu_\B>1$.  Then the results
in \cite{Mu} show that the dual canonical endomorphism for
the inclusion $\A_L^{\max}\otimes \A_R^{\max}\subset \B$ is of
the form $\bigoplus \la \otimes \pi(\la)$, where both
$\A_L^{\max}$ and $\A_R^{\max}$ are local extensions of
$\Vir_c$ and $\la$ runs
through a proper subsystem of the system of
the irreducible DHR endomorphisms of the net $\A_L^{\max}$ and
$\pi$ is an isomorphism from such system  onto another
subsystem of irreducible DHR endomorphisms of the net $\A_R^{\max}$.
Both $\A_L^{\max}$ and $\A_R^{\max}$ are in the
classification list of \cite[Theorem 5.1]{KL}, and now
they are isomorphic.  Recall that at least one of the
two subsystems is a proper subsystem, since $\mu_\B>1$, and the parity
symmetry condition now implies that both subsystems are proper.

First suppose that the map $\pi$ is trivial.
Then the $Q$-system for the inclusion $\A_L^{\max}\otimes
\A_R^{\max}\subset \B$ is
the usual Longo-Rehren $Q$-system arising from the
subsystem by Corollary \ref{coro}, Theorem \ref{vanishing}
and Case 4 of Theorem \ref{SU2-Virasoro}.
Then, Izumi's Galois correspondence \cite[Theorem 2.5]{I3}
shows that we have a further extension $\cC\supset\B$
such that the $Q$-system for
$\A_L^{\max}\otimes \A_R^{\max}\subset \cC$ is the Longo-Rehren
$Q$-system using the entire system of the irreducible DHR endomorphisms
of $\A_L^{\max}$ and the index $[\cC:\B]$ is strictly larger than 1.
We know that the extension $\cC$ arising from the Longo-Rehren
$Q$-system is local.
That is, the net $\B$ is not maximal with respect to local
extensions.

Next suppose that the map $\pi$ is non-trivial.  By checking the
representation categories of the local extensions of the Virasoro
nets classified in \cite[Theorem 5.1]{KL}, we know that only
such non-trivial isomorphisms arise from interchanging of
$2j$ and $4n-2-2j$ of the system $SU(2)_{4n-2}$, where
$j=0,1,\dots,2n-1$, or the well-known
non-trivial automorphism of the system $D_{10}^{\rm even}$.  In both
cases, the map $\pi$ can be extended to an automorphism of the
entire system of irreducible DHR endomorphism of $\A_L^{\max}$ and
we can obtain a proper extension $\cC\supset \B$ in a similar way
to the above case.  Thus, again, the net $\B$ is not maximal with
respect to local extensions.  We summarize these proper
sub-tensor categories of the extensions of the
Virasoro tensor categories $\Vir_c$ ($c<1$)
with trivial or non-trivial
automorphisms as in Table \ref{tab-sub-cat}.  Each entry
``nontrivial'' means that we have a unique nontrivial
automorphism for the sub-tensor category.  For example,
the sub-tensor category $SU(2)_{6}^{\rm even}$ of
$(A_7, A_8)$ appears in the case $n=8$ of the 4th entry
having a trivial automorphism and the case $n=2$ of the
5th entry having a nontrivial automorphism.  We thus have
exactly two non-maximal local conformal nets for this
sub-tensor category.

\begin{table}[htbp]
\begin{center}
\begin{tabular}{|c|c|c|c|}\hline
$m$ & Tensor category & Sub-tensor category & Automorphism \\ \hline
$n$ & $(A_{n-1}, A_n)$ & $\{\id\}$ & trivial \\ \hline
$n$ & $(A_{n-1}, A_n)$ & $\Z/2\Z$ & trivial \\ \hline
$n$ & $(A_{n-1}, A_n)$ & $SU(2)_{n-2}$ & trivial \\ \hline
$n$ & $(A_{n-1}, A_n)$ & $SU(2)_{n-2}^{\rm even}$
& trivial \\ \hline
$4n$ & $(A_{4n-1}, A_{4n})$ & $SU(2)_{4n-2}^{\rm even}$
& nontrivial \\ \hline
$n$ & $(A_{n-1}, A_n)$ & $SU(2)_{n-1}$ & trivial \\ \hline
$4n-1$ & $(A_{4n-2}, A_{4n-1})$ & $SU(2)_{4n-2}^{\rm even}$
& nontrivial \\ \hline
$n$ & $(A_{n-1}, A_n)$ &
$SU(2)_{n-2}^{\rm even}\times SU(2)_{n-1}^{\rm even}$ & trivial \\ \hline
$4n$ & $(A_{4n-1}, A_{4n})$ &
$SU(2)_{4n-2}^{\rm even}\times SU(2)_{4n-1}^{\rm even}$
& nontrivial \\ \hline
$4n-1$ & $(A_{4n-2}, A_{4n-1})$ &
$SU(2)_{4n-3}^{\rm even}\times SU(2)_{4n-2}^{\rm even}$
& nontrivial \\ \hline
$4n+1$ & $(A_{4n}, D_{2n+2})$ & $\{\id\}$
& trivial \\ \hline
$4n+1$ & $(A_{4n}, D_{2n+2})$ & $SU(2)_{4n-1}^{\rm even}$
& trivial \\ \hline
$4n+1$ & $(A_{4n}, D_{2n+2})$ & $D_{2n+2}^{\rm even}$
& trivial \\ \hline
$4n+2$ & $(D_{2n+2}, A_{4n+2})$ & $\{\id\}$
& trivial \\ \hline
$4n+2$ & $(D_{2n+2}, A_{4n+2})$ & $SU(2)_{4n+1}^{\rm even}$
& trivial \\ \hline
$4n+2$ & $(D_{2n+2}, A_{4n+2})$ & $D_{2n+2}^{\rm even}$
& trivial \\ \hline
$11$ & $(A_{10}, E_6)$ & $\{\id\}$ & trivial \\ \hline
$11$ & $(A_{10}, E_6)$ & $\Z/2\Z$ & trivial \\ \hline
$11$ & $(A_{10}, E_6)$ & $SU(2)_2$ & trivial \\ \hline
$11$ & $(A_{10}, E_6)$ & $SU(2)_9^{\rm even}$ & trivial \\ \hline
$11$ & $(A_{10}, E_6)$ & $\Z/2\Z\times SU(2)_9^{\rm even}$
& trivial \\ \hline
$12$ & $(E_6, A_{12})$ & $\{\id\}$ & trivial \\ \hline
$12$ & $(E_6, A_{12})$ & $\Z/2\Z$ & trivial \\ \hline
$12$ & $(E_6, A_{12})$ & $SU(2)_2$ & trivial \\ \hline
$12$ & $(E_6, A_{12})$ & $SU(2)_{11}^{\rm even}$ & trivial \\ \hline
$12$ & $(E_6, A_{12})$ & $\Z/2\Z\times SU(2)_{11}^{\rm even}$
& trivial \\ \hline
$17$ & $(A_{16}, D_{10})$ & $D_{10}^{\rm even}$
& nontrivial \\ \hline
$18$ & $(D_{10}, A_{18})$ & $D_{10}^{\rm even}$
& nontrivial \\ \hline
$29$ & $(A_{28}, E_8)$ & $\{\id\}$
& trivial \\ \hline
$29$ & $(A_{28}, E_8)$ & $SU(2)_3^{\rm even}$
& trivial \\ \hline
$29$ & $(A_{28}, E_8)$ & $SU(2)_{27}^{\rm even}$
& trivial \\ \hline
$30$ & $(E_8, A_{30})$ & $\{\id\}$
&  trivial \\ \hline
$30$ & $(E_8, A_{30})$ & $SU(2)_3^{\rm even}$
&  trivial \\ \hline
$30$ & $(E_8, A_{30})$ & $SU(2)_{29}^{\rm even}$
&  trivial \\ \hline
\end{tabular}
\caption{Proper sub-tensor categories of extensions of the
Virasoro tensor categories $\Vir_c$ with automorphisms}
\label{tab-sub-cat}
\end{center}
\end{table}

Thus we have proved that the net $\B$ with parity symmetry has
$\mu_\B=1$ if and only if it is maximal with respect to local
extensions.  In such a case, we say that $\B$ is a maximal net.
These results, together with Theorem \ref{classif}, imply the
following main theorem of this paper immediately.

\begin{theorem}\label{classif2}
The above correspondence from $\B$ to $Z$ in Theorem \ref{classif}
gives a bijection from
the set of isomorphism classes of such 
maximal two-dimensional nets with parity symmetry and central charge
less than $1$ to
the set of modular invariants $Z$ in Table \ref{CIZ-mod}.
\end{theorem}

Furthermore, the above discussions on the possible proper
sub-tensor categories of the extensions of the
Virasoro tensor categories $\Vir_c$ ($c<1$)
with trivial or non-trivial automorphisms imply that non-maximal
two-dimensional local conformal nets with parity symmetry and
central charge less than $1$ are classified according to
Table \ref{tab-sub-cat}, since we have 2-cohomology vanishing
for all these tensor categories by Theorem \ref{SU2-Virasoro}.

\begin{theorem}
The non-maximal two-dimensional local conformal nets with parity
symmetry and central charge less than $1$ are classified bijectively,
up to isomorphism, according
to the entries in Table \ref{tab-sub-cat}.
\end{theorem}

\medskip

\noindent{\bf Acknowledgments.}
A part of this work was done during a visit of the first-named
author to Universit\`a di Roma ``Tor Vergata''.  Another part was
done while both authors stayed at the 
Mathematisches Forschungsinstitut Oberwolfach
for a miniworkshop ``Index theorems and modularity in operator algebras''.
We thank M. Izumi for useful discussions.
We gratefully acknowledge the support of
GNAMPA-INDAM and MIUR (Italy),
Grants-in-Aid for Scientific Research, JSPS (Japan)
and the Mathematisches Forschungsinstitut Oberwolfach.

\end{document}